\documentclass[a4paper]{cas-sc}

\usepackage[authoryear]{natbib}
\usepackage{graphicx}
\usepackage[utf8]{inputenc}
\usepackage{hyperref}
\usepackage{multicol}
\usepackage{gensymb}
\usepackage{xcolor}
\usepackage{caption}
\usepackage{amsmath}
\usepackage{enumitem}
\usepackage[normalem]{ulem}
\usepackage[
  separate-uncertainty = true,
  multi-part-units = repeat
]{siunitx}
\usepackage{amssymb}
\usepackage{mathtools}
\usepackage{makecell}

\def\tsc#1{\csdef{#1}{\textsc{\lowercase{#1}}\xspace}}
\tsc{WGM}
\tsc{QE}
\tsc{EP}
\tsc{PMS}
\tsc{BEC}
\tsc{DE}

\newcommand{\Lagr}{\mathcal{L}}
\newcommand{\Z}{\mathcal{Z}}
\newcommand{\B}{\mathcal{B}}

\renewcommand*{\thefootnote}{\fnsymbol{footnote}}

\begin{document}
\let\WriteBookmarks\relax
\def\floatpagepagefraction{1}
\def\textpagefraction{.001}
\shorttitle{Bayesian analysis of \textit{Juno}/JIRAM's NIR observations of Europa}
\shortauthors{Ishan Mishra}

\title[mode = title]{Bayesian analysis of \textit{Juno}/JIRAM's NIR observations of Europa}                     



\author[1,2]{Ishan Mishra}[orcid=0000-0001-6092-7674]
\cormark[1]
\ead{im356@cornell.edu}

\author[1,2]{Nikole Lewis}[orcid=0000-0002-8507-1304]

\author[1,2]{Jonathan Lunine}[orcid=0000-0003-2279-4131]

\author[1]{Paul Helfenstein}[orcid=0000-0003-4870-1300]

\author[1,2]{Ryan J. MacDonald}[orcid=0000-0003-4816-3469]

\author[3]{Gianrico Filacchione}[orcid=0000-0001-9567-0055]

\author[3]{Mauro Ciarniello}[orcid=0000-0002-7498-5207]

\address[1]{Department of Astronomy, Cornell University, 122 Sciences Drive, Ithaca, NY 14853, USA}

\address[2]{Carl Sagan Institute, Cornell University, 122 Sciences Drive, Ithaca, NY 14853, USA}

\address[3]{INAF-IAPS, Istituto di Astrofisica e Planetologia Spaziali, Area di Ricerca di Tor Vergata, via del Fosso del Cavaliere, 100, 00133, Rome, Italy}


\begin{abstract}
Decades of observations of Europa's surface in the near-infrared (NIR), spanning spacecrafts like Galileo, Cassini and New Horizons, along with ground based observations, have revealed a rich mixture of species on Europa's surface. Adding to the NIR data of Europa, Juno spacecraft's spectrometer JIRAM has observed it in the 2-5 $\mu$m wavelength region. Here we present analysis of select spectra from this dataset, focusing on the two forms of water-ice - amorphous and crystalline. We were limited in our ability to include other dominant Europan species, like acid hydrate, due to unavailability of their optical constants over the entire JIRAM wavelength range. We also take this as an opportunity to present a novel Bayesian spectral inversion framework. Traditional spectral fitting methods, for example a grid-based search of parameter space, lack a systematic way to quantify detection significances of the species included in the model, statistically constrain surface properties and explore degenerecies of solution. Our Bayesian inference framework overcomes these shortcomings by confidently detecting amorphous and crystalline ice in the JIRAM data and permits probabilistic constraints on their compositions and average grain sizes to be obtained. We first validate our analysis framework using simulated spectra of amorphous and crystalline ice mixtures and a laboratory spectrum of crystalline ice. We next analyze the JIRAM data and, through Bayesian model comparisons, find that a two-component, intimately mixed model of amorphous and crystalline ice, henceforth referred to as TC-IM, is strongly preferred (at $26 \sigma$ confidence) over a two-component model of the same materials but where their spectra are areally/linearly mixed. We also find that the TC-IM model is strongly preferred (at > $30\sigma$ confidence) over single-component models with only amorphous or crystalline ice, indicating the presence of both these phases of water ice in the data. Given the high SNR of the JIRAM data, abundances and grain sizes of amorphous and crystalline ice are very tightly constrained for the analysis with the TC-IM model. The solution corresponds to a mixture with a very large number density fraction (99.952$_{-0.001}^{+0.001}$ \%) of small (23.12$_{-1.01}^{+1.01}$ microns) amorphous ice grains, and a very small fraction (0.048$_{-0.001}^{+0.001}$ \%) of large (565.34$_{-1.01}^{+1.01}$ microns) crystalline ice grains. The overabundance of small amorphous ice grains we find is consistent with previous studies. The maximum-likelihood spectrum of the TC-IM model, however, is in tension with the data in the regions around 2.5 and 3.6 $\mu$m, and indicates the presence of non-ice components not currently included in our model. Our new technique therefore holds the promise of being able to identify these minor species hiding in Europan reflectance data in future work and constrain their abundances and physical properties.
\end{abstract}

 
\maketitle

\section{Introduction}

Constraining the surface composition of Jupiter's icy satellite Europa is critical for determining the habitability of its subsurface ocean. Materials from the interior ocean might have been emplaced on Europa's surface through vertical movement of warm ice or fluids \citep[e.g.][]{kattenhorn_evidence_2014,blaney_mapping_2017}. Apart from the dominant water ice, various other components like hydrated sulfuric acid \citep{carlson_sulfuric_1999, carlson_distribution_2005}, hydrated sulfates \citep{mccord_salts_1998, dalton_linear_2007}, chlorinates \citep{fischer_spatially_2015, ligier_vlt/sinfoni_2016, trumbo_sodium_2019} and oxidants \citep{hansen_widespread_2008, hand_keck_2013} have been tentatively identified in the reflectance spectroscopy data of Europa. The Near-InfraRed or NIR wavelength regime ($\sim 1-5$ $\mu$m) has been especially fruitful in disentangling the spectral signature of non-water-ice components on Europa's surface, as the compositionally dominant water-ice's reflectance tapers off with increasing wavelength \citep[e.g.][]{dalton_linear_2007}. 


Given the overall spectral properties of Europa's surface, crystalline and amorphous water ice are the dominant species one would expect. Going back to Galileo, we have evidence that Europa's surface has both these water ice phases \citep{hansen_amorphous_2004,carlson_europas_2009}. \citet{filacchione_serendipitous_2019} used a spectral-indices based analysis on the \textit{Juno}/JIRAM observations of Europa to detect both forms of water-ice. Within the range of temperatures observed on Europa's surface ($\sim 80-130$ K \citep{spencer_temperatures_1999}), one would expect that all amorphous ice would be transformed into hexagonal crystalline ice in less than 20 years. However, mechanisms like condensation of sublimated and sputtered molecules and irradiation from UV, electrons and ions (which are plentiful in Jupiter's intense magnetospheric environment), can lead to amorphization of water-ice. 

Adding to the abundant spacecraft NIR data of Europa's surface from missions like Galileo \citep{carlson_near-infrared_1996}, Cassini \citep{mccord_cassini_2004} and New Horizons \citep{grundy_new_2007}, Juno's near-infrared (NIR) spectrometer JIRAM, the Jovian InfraRed Auroral Mapper \citep{adriani_jiram_2017}, has obtained several serendipitous spectra of Europa in the $2-5 \ \mu$m wavelength range \citep{filacchione_serendipitous_2019}. In this work, we present a water-ice composition analysis of selected Juno/JIRAM spectra using an analytical bidirectional reflectance model \citep{hapke_bidirectional_1981,hapke_theory_2012}  wrapped in a Bayesian inference framework. 

Bayesian inference overcomes many of the pitfalls of traditional fitting methods when working with non-linear models \citep{andrae_dos_2010}. In the `classical' forward modelling approach to fitting data, one compares a few models to the data to look for a single best-fitting solution in a grid-based search. In contrast, the Bayesian inversion method takes the data as the starting point and uses a statistical sampling algorithm to generate millions of models to systematically explore the range of parameters compatible with the data. The overall advantages of using a Bayesian inference framework can be summarized as 1) it provides the ability to encode prior information about parameters in the model 2) it allows exploration of a large parameter space to look for non-unique solutions and obtain statistical constraints on the underlying model parameters (e.g. grain size) 3) it is robust to low-SNR data when it comes to constraining the parameters \citep{trotta_bayesian_2017}, as can be demonstrated through analysis on simulated data (see section \ref{sec:synthetic_and_lab}), 3) Bayesian model comparisons provide a mechanism to quantitatively compare different conceptual models \citep{trotta_bayesian_2017}, for example, linearly v/s intimately mixed compositional endmembers. 

Bayesian inference for inverse problems have a rich history in planetary science, specifically in planetary geophysics \citep[e.g.][]{tarantola_inverse_1982,stark_mercurys_2015,cornwall_planetary_2016}, atmospheric remote sensing \citep[e.g.][]{irwin_nemesis_2008,koukouli_water_2005,nixon_meridional_2007}, exoplanet detection \citep{ford_quantifying_2005} and exoplanet atmospheric studies \citep[e.g.][]{line_systematic_2013,lee_atmospheric_2013, madhusudhan_atmospheric_2018}. However, Bayesian methods have been underutilized in planetary surface reflectance spectroscopy, with a limited number of studies that have applied a Bayesian Markov Chain Monte Carlo (MCMC) approach to the inversion problem \citep[e.g.][]{fernando_surface_2013, schmidt_realistic_2015,  fernando_martian_2016, lapotre_probabilistic_2017, lapotre_compositional_2017,rampe_sand_2018,belgacem_regional_2020}. As for icy satellites, a recent application is the study of Europa’s photometric - but not spectroscopic - data using Bayesian inference by \citet{belgacem_regional_2020}.

Here, we present a new Bayesian approach to the analysis of Europa surface spectra. We illustrate our technique by conducting a preliminary analysis focusing on two endmembers - amorphous and crystalline water-ice - and demonstrate its ability to directly constrain the abundance and grain-sizes. Although Europa's surface is also dominated by heavily hydrated species, whose total abundance sometimes far exceeds that of water ice in certain areas, we don't include such species in our analysis due to the lack of availibility of their optical constants over the entire JIRAM wavelength range (2-5 $\mu$m).  We describe our methodology in section \ref{sec:methodology}, starting with the data and its pre-processing in section \ref{sec:data_preprocessing}. Section \ref{sec:data_analysis} details the forward model of our analysis framework (the Hapke bidirectional reflectance model) and the Bayesian inference framework. In section \ref{sec:synthetic_and_lab}, we illustrate the use of the Bayesian framework on simulated/synthetic reflectance data of an amorphous and crystalline ice mixture, followed by an application to laboratory reflectance spectra of crystalline water ice. Having illustrated the efficacy of this framework in section \ref{sec:synthetic_and_lab}, we show its application to the \textit{Juno}/JIRAM data in section \ref{sec:jiram}, followed by the discussion of the results and conclusions in section \ref{sec:disc_and_conc}.

\section{Methodology} \label{sec:methodology}

\subsection{Data pre-processing} \label{sec:data_preprocessing}

Juno's near-IR spectrometer JIRAM observed Europa's surface during  orbit numbers $2$,$8$,$9$ and $11$ which occured between 2016 and 2018. The complete dataset and the associated observation geometry parameters are shown in Table \ref{Tab: data} of \citet{filacchione_serendipitous_2019}. We obtained the radiance factor ($\textrm{I/F}$) spectrogram (normalized at 2.227 $\mu$m) of the observations through personal communication with the authors (in sections \ref{sec:synthetic_data} and \ref{sec:lab} we show examples where the results are not affected by normalization of the data, for the parameters considered in this work).  It should be noted that the retrieval of JIRAM's FOV footprint on Europa's surface has uncertainties, detailed in section 4 of \citet{filacchione_serendipitous_2019}. Exact locations on Europa's surface of the spectrometer's pixels and the corresponding incidence and emission angles are not available; instead we have a range of incidence and emission angles for each data set corresponding to an observational sequence. Hence, in this work we focus on observational datasets from sequences where the range of incidence and emission angles is small, so that the angles can be approximated with mean values. As we are interested in studying the reflected sunlight off Europa's surface, we focus on incident and emission angles less than 90$\degree$ to avoid contributions from Jupiter-shine. With these criteria in mind, we choose to focus on the four spectra of observation session JM0081\_170901\_105708, described in Table \ref{Tab: data}.

\begin{table}
\captionsetup{width=\textwidth}
\caption{Table \ref{Tab: data}: The JIRAM observation session used in this work and its geometry parameters. The longitude coordinates are West from the Europan sub-Jupiter point and the latitude coordinates are North from the Europan equator. For a complete map of regions mapped by JIRAM on Europa, see Figure 1 of \citet{filacchione_serendipitous_2019}.}
\centering
\resizebox{\textwidth}{!}{%
\begin{tabular}{ c c c c c c c c} 
 \hline
 \textbf{Session} &	\textbf{Latitude (deg)} & \textbf{Longitude (deg)} & \textbf{Incidence (deg)} & \textbf{Emission (deg)}	& \textbf{Phase (deg)} & \textbf{Resolution} & \textbf{\# spectra}\\ 
 \hline
 \textbf{Orbit-Date-Time} &	\textbf{Min  \ \ Max} &	\textbf{Min \ \	Max} &	\textbf{Min \ \ Max} &	\textbf{Min \ \ Max} &	\textbf{Min \ \ Max} &	\textbf{(km/pix)} &  \\
 \hline
 JM0081\_170901\_105708 &	20.6 \ \ 24.0 &	37.4 \ \	40.8 &	28.5 \ \	28.5 &	73.4 \ \	73.4 &	91.5	\ \ 91.5 &	88 &	4 \\
 \hline
\end{tabular}}
\label{Tab: data}
\end{table}

For each extracted and normalized (at 2.227 $\mu$m) $\rm I/F$ spectrum we stepped through the following post-processing procedure:

\begin{enumerate}
    \item Remove NANs.
    \item Remove the data in the $3.7$ and $3.8$ $\mu$m region as they have uncorrectable systematic errors caused by the instrumental filter-order sorting interface. It is seen as a dark vertical strip at $3.8$ $\mu$m in Figure~ 2 of \citet{filacchione_serendipitous_2019}. 
    \item De-spike or remove outliers in the data. We do this by moving a 20-channel boxcar window across the data array (the data spans 336 wavelength channels) and flagging points. A point is flagged if it is more than $\sim 4\sigma$ away from the median of the data values within the window. The window length and parameters for the threshold were chosen experimentally to ensure that no more than $\sim 5\%$ of the total number of the data points were discarded. 
\end{enumerate}

Once the data were post-processed, we proceeded to calculate the error bars/noise in the (I/F) spectra. The spectrogram we retrieved from \citet{filacchione_serendipitous_2019} had been corrected for non-linear readout noise (high frequency noise between odd and even spectral bands) introduced by the detector's multiplexer \citep{filacchione_saturns_2007}. We proceeded to calculate the instrumental noise affecting JIRAM data, which is computed by comparing the measured spectral radiance with respect to the noise-equivalent spectral radiance (NESR). For a single (I/F) spectrum, the noise is calculated as

\begin{gather} \label{eq:IF_noise}
\dfrac{\textrm{I}}{\textrm{F}}(\lambda)_{noise} = \dfrac{\dfrac{\textrm{I}}{\textrm{F}}(\lambda)}{\textrm{SNR}(\lambda)}
\end{gather}
where SNR is the signal-to-noise ratio calculated for that spectrum. The SNR for an individual spectrum is simply defined as the ratio of the equivalent spectral radiance $R$ to the NESR

\begin{gather}
     \textrm{SNR}(\lambda) =  \left(\dfrac{\textrm{R}(\lambda)}{\textrm{NESR}(\lambda)}\right)
\end{gather}
For a given (I/F) spectrum, we calculate the equivalent spectral radiance R($\lambda$) using the formula

\begin{gather}
    \dfrac{\textrm{I}}{\textrm{F}}(\lambda) = \frac{4\pi D^2 \textrm{R}(\lambda)}{\textrm{SI}(\lambda)} \\
    \implies \textrm{R}(\lambda) = \dfrac{\dfrac{\textrm{I}}{\textrm{F}}(\lambda) \ {\textrm{SI}(\lambda)} }{4\pi D^2}
\end{gather}
where $\textrm{SI}(\lambda)$ is the solar irradiance measured at 1 AU \citep{kurucz_including_2009} and $D$ is Europa's heliocentric distance (in AU) at the time of observation. The NESR was obtained through personal communication with the authors of \citet{filacchione_serendipitous_2019} (see Figure~ 3 therein). The NESR gives the minimum spectral radiance in W m$^{-2}$ $\mu$m$^{-1}$ sr$^{-1}$ corresponding to 1 DN (digital number) as measured by the instrument at operative temperature conditions. It is defined as 

\begin{gather}
    \textrm{NESR}(\lambda) = \dfrac{\textrm{Std. deviation of }(DC(\lambda) + B(\lambda))}{\textrm{Resp}(\lambda)}
\end{gather}
where $DC(\lambda)$ is the dark current, $B(\lambda)$ is the background signal equivalent to the sky and Resp$(\lambda)$ is the instrument responsivity at wavelength $\lambda$. 

\begin{figure}[pos=htbp!]
    \centering
    \includegraphics[width=0.5\linewidth]{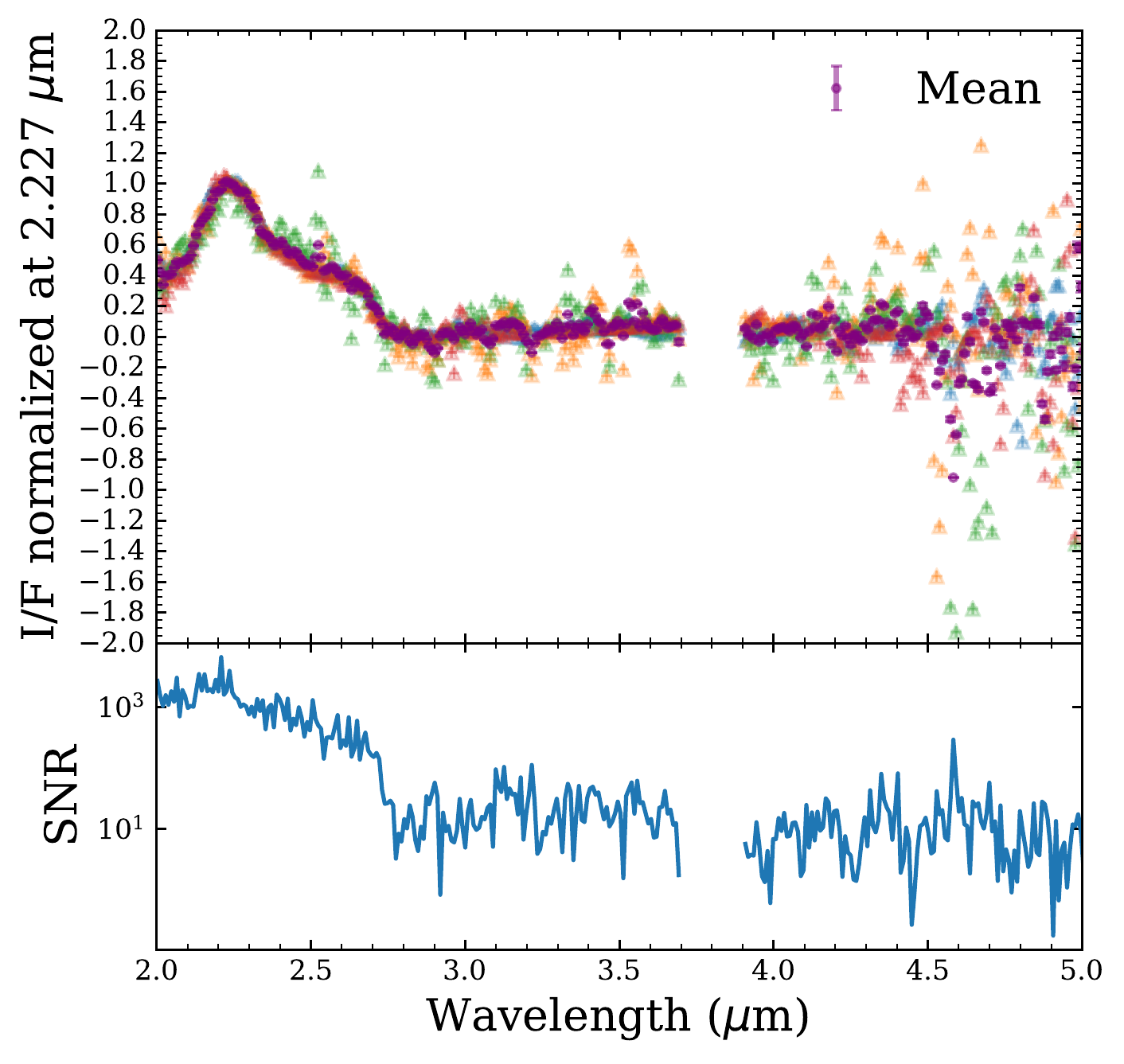}
    \caption{Top: JM0081\_170901\_105708 observations (triangles colored in red, green, yellow and blue for the 4 spectra in the dataset) and their mean (purple circles). Bottom: The signal-to-noise ratio corresponding to the mean data and the RMS noise.}
    \label{fig:data}
\end{figure}

Since all four of the spectra have similar observation geometries and correspond to geographically close regions (see Figure 1. of \citep{filacchione_serendipitous_2019}), and to simplify the analysis, in this work we use the mean spectrum of the \\ JM0081\_170901\_105708 observation.  For a set of $N$ (I/F) spectra of a given observational sequence, the root-mean-square noise is calculated as

\begin{gather} \label{eq:rms_noise}
    \dfrac{\textrm{I}}{\textrm{F}}(\lambda)_{rms \ noise} = \sqrt{\dfrac{1}{N}\sum_{i=1}^{N} \Big(\dfrac{\textrm{I}}{\textrm{F}}(\lambda)_{i,noise}\Big)^2}
\end{gather}
%
%
Here, (I/F)$_{noise}$ is the noise corresponding to a single (I/F) spectrum (eq. \ref{eq:IF_noise}). We recognize that our errors are likely underestimated due to normalization factor and other unknown sources. The four JM0081\_170901\_105708 observations, their mean spectrum and the signal-to-noise ratio (SNR) corresponding to the RMS error are shown in Figure~ \ref{fig:data}. The analysis methodology of this data with a Bayesian inference framework is detailed in what follows.

\subsection{Data Analysis} \label{sec:data_analysis}

\begin{figure}[pos=htbp!]
    \centering
    \includegraphics[width=0.75\linewidth]{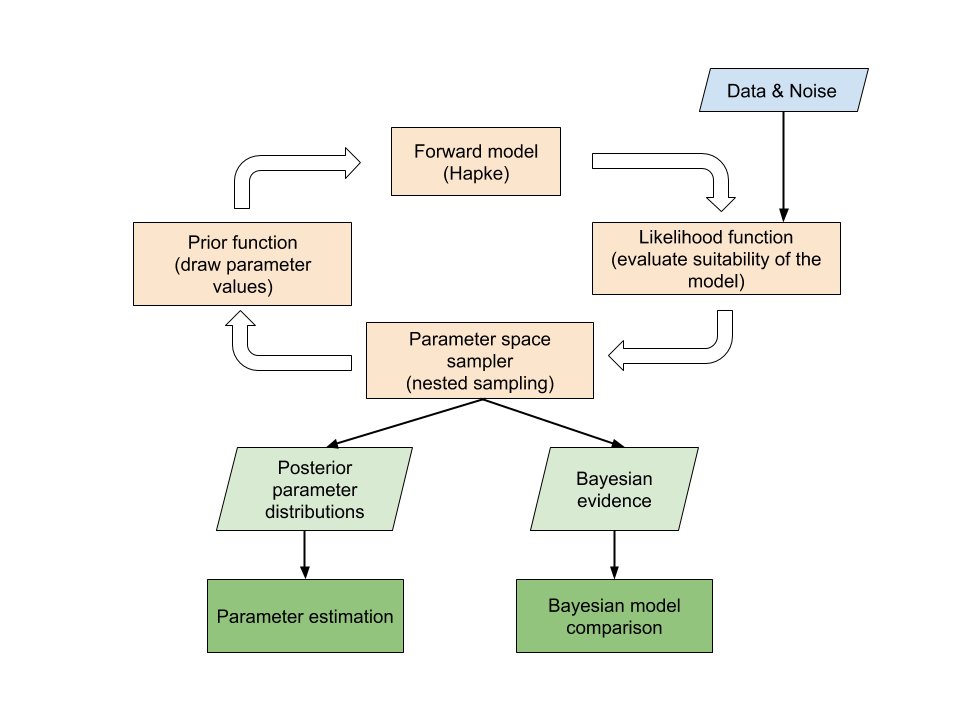}
    \caption{Workflow for our Bayesian retrieval framework. The blue box containing `Data \& Noise` represents the inputs to this framework, specifically to the likelihood function \ref{eq:bayes_theorem}. The orange boxes are part of the iterative process that samples the posterior function. The posterior function encodes the probability of a set of model parameters in light of the data. A sampling algorithm like MCMC or nested sampling (used in this work) efficiently samples the multi-dimensional model parameter space, whose bounds are defined by the prior function, to approximate the posterior probability function. The outputs of this sampling process and their corresponding uses are shown in light-green and dark-green boxes, respectively. Firstly, the list of multi-dimensional samples can be marginalized to get a probability distribution for each parameter and estimate metrics like confidence intervals. Secondly, in some cases, we can also obtain the Bayesian evidence (eq. \ref{eq:bayes_theorem}) of the model,  which quantifies its overall goodness-of-fit and is used for model comparison.}
    \label{fig:workflow}
\end{figure}

In this section, we detail the various moving parts of the Bayesian inference, also known as `retrieval framework'. Our workflow for its application to reflectance data is summarized in section \ref{sec:workflow_overview} and Figure~ \ref{fig:workflow}. Next, in section \ref{sec:forward_model} we describe the Hapke radiative transfer model that serves as the forward model in our retrieval framework, followed by an overview of the other components of the retrieval framework in section \ref{sec:bayesian_framework}.

\subsubsection{Workflow Overview} \label{sec:workflow_overview}

As noted by \citet{lapotre_compositional_2017}, the process of interpreting a reflectance spectrum of planetary surface regolith broadly requires:

\begin{enumerate}
    \item Information about the observational geometry, i.e., the incidence, emission and phase angles.
    \item A forward model that outputs a reflectance spectrum given the observational geometry parameters and information (optical constants, average grain size, etc.) about the components of the mixture. 
    \item A methodology/framework to infer compositional information by inverting the data through repeated calls to the forward model. 
\end{enumerate}

The forward model we use in this work is a radiative transfer model developed by \citet{hapke_bidirectional_1981, hapke_bidirectional_2012}. It relates the reflectance of a mixture to the linear combination of the single-scattering albedos (if the particles are intimately mixed) or reflectances (if the particles are areally mixed, i.e., distributed in distinct patches) of its constituent endmembers. Since the single-scattering albedo of a material is a function of its optical constants, i.e., the real and imaginary part of the refractive index (\textit{n} and \textit{k} respectively), we need to have the laboratory measured optical constants as a function of wavelength for each endmember we wish to include in our model. In case of water-ice dominated surfaces such as Europa, clues for other endmembers can be found by looking at the distortion in water-ice absorption bands. For Europa specifically, these have been proposed to be caused by hydrated salts and sulfuric acid  \citep{carlson_europas_2009}. In this work however, we are focusing only on constraining amorphous and crystalline water ice.

A detailed discussion of the Bayesian retrieval framework is provided in section \ref{sec:bayesian_framework}. The flowchart in Figure~~\ref{fig:workflow} provides a quick overview and highlights the different parts of a Bayesian retrieval framework at different stages. At the heart of it all is a likelihood function (eq. \ref{eq:bayes_theorem} that calculates the probability of a set of parameters using the model spectrum generated using these parameters. If the data-errors are Gaussian, then this probability is directly related to the chi-squared value of the fit of the model spectrum to the data. The probability values of models generated from thousands of sets of parameters or samples lead to a probability `surface'. To efficiently map/sample this probability surface, we have various tools/algorithms like MCMC (Markov-Chain Monte Carlo) and nested sampling (used in this work) available at our disposal.  The output samples, or collection of parameter values, of a sampling algorithm can be used to evaluate useful descriptors like the maximum-a-posteriori model and posterior distributions of individual model parameters. The Bayesian formulation uses a prior function, which encodes any prior knowledge of the model parameters we are fitting for. If the prior distribution of the parameters is uniform \citep{hogg_data_2010}, which is the case in all the analyses presented in this paper, then the maximum-a-posteriori parameter estimates are identical to the maximum-likelihood estimates . The posterior distribution of a parameter can be used to estimate the uncertainty, typically the 68\% confidence or $\pm 1\sigma$ interval, around its median value. We also get the correlations between different parameters, which sheds light on the degeneracies within the model and non-uniqueness of the high probability-density regions in the parameter-space. Another useful output of a sampling algorithm is the Bayesian evidence or the model evidence, which is the marginal likelihood of the data over the entire parameter space (see eq. \ref{eq:bayes_theorem}), or in other words, the probability of the data given a physical model. This quantity can be directly used for comparing physical models, where a `better' model would have a higher Bayesian evidence value.

\subsubsection{The Forward model} \label{sec:forward_model}

\begin{figure}[pos=htbp!]
    \centering
    \includegraphics[width=0.5\linewidth]{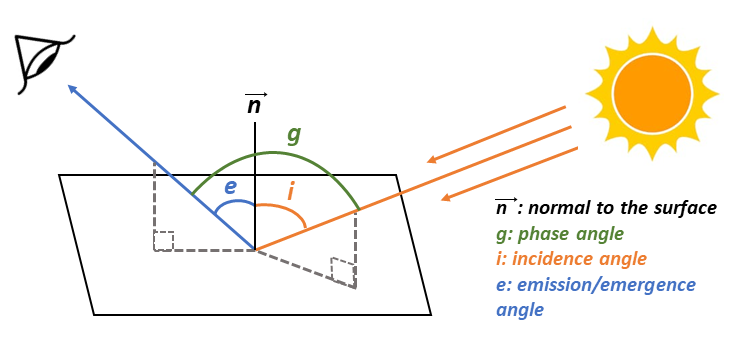}
    \caption{The viewing geometry used by the Hapke model and \textit{Juno}. Adapted from \citet{belgacem_regional_2020}.}
    \label{fig:hapke_geo}
\end{figure}

\citet{hapke_bidirectional_1981,hapke_theory_2012} provided a scheme to calculate the bidirectional reflectance of a surface consisting of particles of arbitrary shape in close proximity to one another, in the geometric optics regime, and for the viewing geometry shown in Figure~\ref{fig:hapke_geo}. This bidirectional reflectance, as a function of wavelength $\lambda$, is given to a good approximation by (eq. 10.5 in \citet{hapke_theory_2012}):

\begin{gather} \label{eq:hapke_RT}
   \dfrac{I}{F}(\mu,\mu_0,g, \lambda) = K \dfrac{\omega(\lambda)}{4}\dfrac{\mu_0}{(\mu + \mu_0)}[P(g,\lambda) + H(\omega,\mu/K)H(\omega,\mu_0/K) - 1]
\end{gather}
Here 
 
 \begin{itemize}
    \item $I/F$ is the radiance factor, which is the ratio of bidirectional reflectance of a surface to that of a perfectly diffuse surface (Lambertian) illuminated at $i=0$. 
    \item $\mu$ is the cosine of the emission angle $e$.
    \item $\mu_0$ is the cosine of the incidence angle $i$.
    \item $g$ is the phase angle.
    \item $K$ is the porosity coefficient.
    \item $\omega$ is the single scattering albedo.
    \item $P$ is the particle phase function.
    \item $H$ is the Ambartsumian-Chandrasekhar function that accounts for multiple scattered component of the reflection \citep{chandrasekhar_radiative_1960}.
 \end{itemize}

Since the phase angle of all of Juno’s observations is $\approx 90 \deg$, due to its polar orbit, we have ignored functions that account for backscattering and other opposition effects which become important at small phase angles. We are also ignoring the photometric effects of large-scale roughness, as in general, moderate topographic slopes have little effect on normalized spectra \citep{hapke_theory_2012}. The major parameters of the models are described in detail in the Appendix.\\

\noindent \emph{Mixing equations} \\
 
\noindent In planetary surface reflectance spectroscopy studies, there are two widely used mixing modalities of interest: areal and intimate \citep{hapke_theory_2012}. In an areal mixture, the surface area viewed by the spectrometer consists of several unresolved, smaller patches, each of which consists of a pure material. The total reflectance of the area in this case is simply a linear sum of each reflectance weighted by area. Hence, this mixing model is known as \textit{linear mixing model} and the total reflectance is given by

\begin{gather} \label{eq:lm_eqn}
    r = \sum_i F_j r_j
\end{gather}
where $r$ is a type of reflectance (the radiance factor in our case), $r_j$ is the same type of reflectance of the $j$th area patch, and $F_j$ is the fraction of area viewed by the detector occupied by the j-th area. 

\noindent On the other hand, in an intimate mixture, different types of particles are mixed homogenously together in close proximity. The averaging process in this \textit{intimate mixing model} is over the individual particle, and certain parameters of the Hapke radiative transfer equation (eq. \ref{eq:hapke_RT}) are volume averages of the different materials in the mixture, weighted by their cross-sectional area. Among the parameters of the model we have described so far, the single scattering albedo $\omega$ and the particle phase function $p$ undergo the averaging process. Assuming that the particles are equant, the volume average single scattering is

\begin{gather} \label{eq:im_eqn}
    \omega_{mix} = \dfrac{\sum_j N_j \sigma_j Q_{Ej} \omega_j}{\sum_j N_j \sigma_j Q_{Ej}} = \dfrac{\sum_j f_j \sigma_j Q_{Ej} \omega_j}{\sum_j f_j \sigma_j Q_{Ej}}
\end{gather}
and the volume average phase function is

\begin{gather}
    p_{mix} = \dfrac{\sum_j N_j \sigma_j Q_{Ej} \omega_j p_j}{\sum_j N_j \sigma_j Q_{Ej} \omega_j} = \dfrac{\sum_j f_j \sigma_j Q_{Ej} \omega_j p_j}{\sum_j f_j \sigma_j Q_{Ej} \omega_j}
\end{gather}

where $N_j$ is the number of particles of type $j$ per unit volume, $\sigma_j (= \pi D^2/4)$ is the cross-sectional area of a particle of type $j$, $Q_{E,j}$ is the volume average extinction efficiency and $\omega_j$ is the single scattering albedo of a particle of type $j$. In the right-most expression of eq. \ref{eq:im_eqn}, the number density $N_j$ has been converted to a number density fraction $f_j(=N_j/\sum N_j)$. The number density $N_j$ can be directly converted to a `mass-density' $M_j$, the mass of particles of type $j$ per unit volume of the mixture, using

\begin{gather} 
    M_j = \dfrac{2}{3} N_j \sigma_j \rho_j D_j
\end{gather}
where $\rho_j$ is the solid density of material $j$ and $D_j$ is the average particle diameter. This relation will be useful later to convert the fitting solutions for $f_j$ and $D_j$, the two main free parameters in our Bayesian framework, into a mass-density fraction $mf_j$, which can be calculated as

\begin{gather} \label{eq:mass_frac}
    mf_j = \dfrac{M_j}{\sum_j M_j} = \dfrac{\dfrac{2}{3} N_j \sigma_j \rho_j D_j}{\sum_j \dfrac{2}{3} N_j \sigma_j \rho_j D_j} = \dfrac{\dfrac{2}{3} f_j \sigma_j \rho_j D_j}{\sum_j \dfrac{2}{3} f_j \sigma_j \rho_j D_j}
\end{gather}

\noindent Although this multi-component reflectance model involves many equations and parameters, the strongest dependence comes from the mixture's single-scattering albedo $\omega_{mix}$ and hence on the individual members' single-scattering albedos. Equations \ref{eq:ssa}-\ref{eq:<D>} show us that for a given component, $\omega$ depends on its optical constants and average grain size. For the two components in our model, amorphous and crystalline ice, we have used optical constants measured at 120 K and published by \citet{mastrapa_optical_2009}.

\subsubsection{Bayesian inference framework} \label{sec:bayesian_framework}

Equipped with the optical constants of the two components in the mixture, and assuming that the observational geometry parameters are known and have been fixed, we let the following be free parameters for the fitting exercise:

\begin{enumerate}
    \item grain-size or diameter $D_i$ of each mixture component
    \item number density fraction $f_i$ of each member component (needed if there is more than one component in the model)
    \item filling factor $\phi$ (see eq. \ref{eq:K_defn})
\end{enumerate}

\begin{figure}[pos=htbp!]
    \centering
    \includegraphics[width=0.5\linewidth]{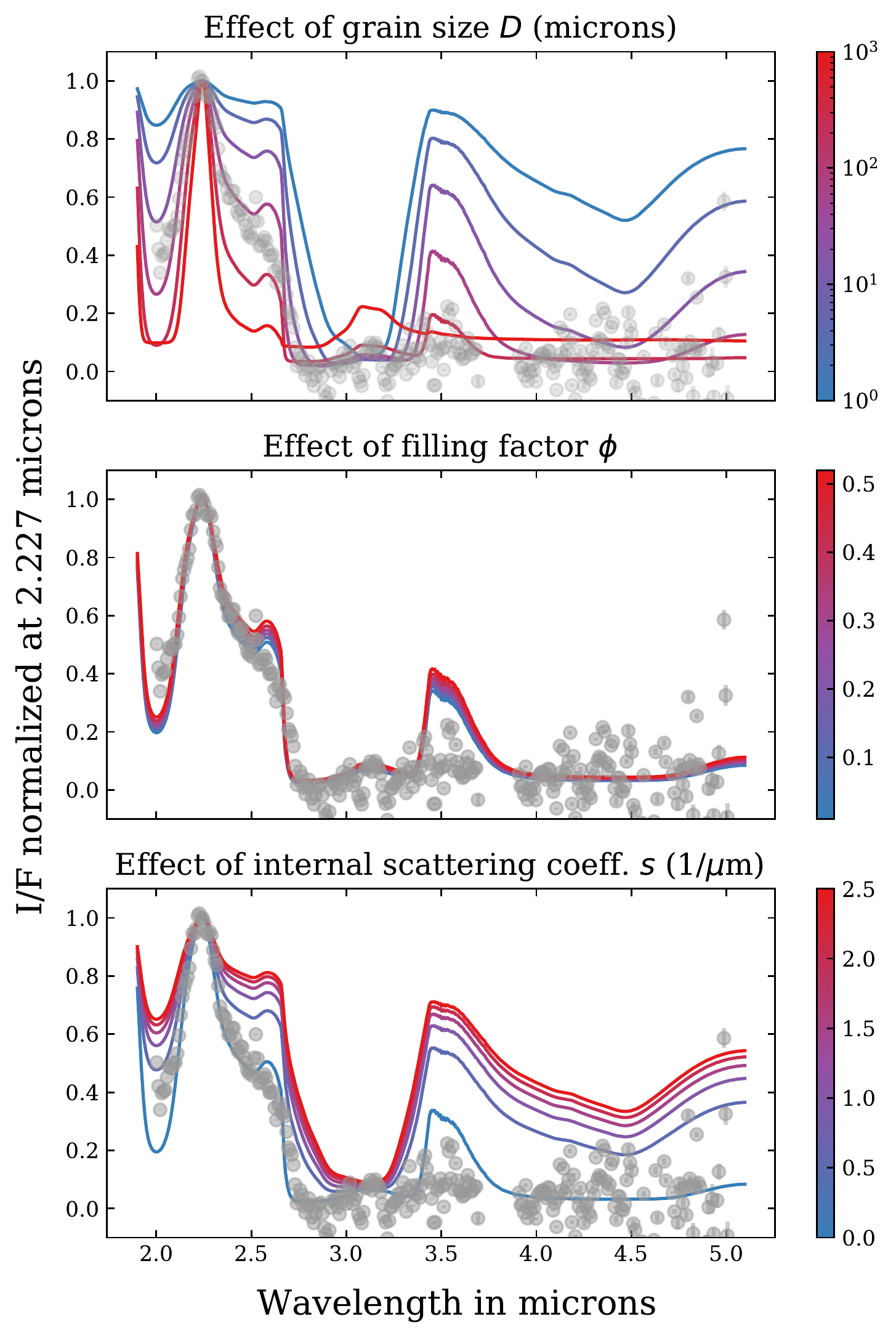}
    \caption{Sensitivity of model amorphous ice spectrum (at $T=120$K) for different values of grain size $D$ (top), filling factor $\phi$ (middle) and internal scattering coefficient $s$ (bottom), plotted along with the JIRAM data. Grain size $D$ has been fixed to 100 microns for the cases of varying $\phi$ and $s$. For cases with varying $D$ and $s$, filling factor $\phi$ has been set to 0.001 (such that the porosity coefficient is  $\sim1.0$ (see eq. \ref{eq:K_defn})). For cases with varying $D$ and $\phi$, $s$ has been set to 0. The observation geometry parameters (incidence, emission and phase angles) are same as the data (Table \ref{Tab: data}).}
    \label{fig:sensitivity}
\end{figure}

The effects of grain size $D$ and the filling factor $\phi$ on a normalized model spectrum of amorphous ice are shown in the top and middle panels of Figure~ \ref{fig:sensitivity}. The effect of $D$ is apparent, with increasing grain-size causing more absorption and hence reduced reflectance features. On the other hand, changing the filling factor $\phi$ (within its permissible range) does not alter the normalized spectrum very much. However, we still set $\phi$ as a free parameter as wish to test our analysis framework's ability to constrain it. 

Another choice for a free parameter could have been the internal scattering coefficient $s$ (appears in eq. \ref{eq:in_trans_factor}), which characterizes the density of scatterers within the particle. 
However, as the bottom panel of Figure~ \ref{fig:sensitivity} shows, even small values of $s$ makes a reasonably well-fitting model spectrum, with $s$ at its default lowest value of 0, severely stray away from the JIRAM data. Hence, we deemed that its not a necessary parameter to fit for and have fixed it to be equal to 0. This also helps us avoid degenerate solutions, since the effect of increasing $s$ looks qualitatively similar to decreasing the grain size $D$ in Figure \ref{fig:sensitivity}.

The free parameters are basically the `knobs' of the model that can be turned to change the output spectrum and explore how the fit to the data changes. We carry out this parameter-space exploration using a Bayesian framework. Moreover, since we can work with different permutations of the two components in our model (amorphous and crystalline ice), we must assess the feasibility of one model over another model in explaining the data. For example, we might want to compare the suitability of a one-component model with only amorphous ice over a two-component model with both amorphous and crystalline ice, or vice-versa, to fit the data. This goal can be accomplished with Bayesian model comparison. Both these goals are the end products of our Bayesian inference framework as shown in Figure~ \ref{fig:workflow} and elucidated here. \\ 

\noindent \emph{Bayes' Theorem} \\

\noindent Following the scheme of \citet{macdonald_hd_2017}, we consider a forward model $M$ which is described by a set of variable parameters denoted by $\theta_j$, where $j$ is the dimension index of the parameter space. In our case, $M$ would be the Hapke reflectance model and $\theta_j$ would be the abundances, grain-sizes of the components, etc.. Our \textit{a priori} expectations of the parameters are encoded in the \textit{prior probability density function $p(\theta_j|M)$}. Using a set of observations $y_{obs}$, which in our case are the JIRAM I/F data points shown in Figure~\ref{fig:data}, we can formally update our knowledge of the values of these parameters via \emph{Bayes' theorem}

\begin{gather} \label{eq:bayes_theorem}
    p(\theta_j|y_{obs},M) = \dfrac{p(y_{obs}|\theta_j,M) p(\theta_j|M)}{\int p(y_{obs}|\theta_j,M) p(\theta_j|M) d\theta_j} \equiv \dfrac{\Lagr(p(y_{obs}|\theta_j,M)) \pi (\theta_j|M)}{\Z(y_{obs}|M)}
\end{gather}
where $\Lagr$ is the conventional notation for the likelihood function, $\pi$ is the prior and $\Z$ is the Bayesian evidence. \\ $p(\theta|y_{obs},M)$ is the posterior function which gives us the probability of the model parameters in light of the new data and our prior knowledge of the parameters. 

For a model $M_i$, the likelihood function, $\Lagr (y_{obs}| \theta, M_i)$, expresses how likely the data is given the model and its parameters. If the noise on the data points can be assumed to be Gaussian and independent, the likelihood is given by

\begin{gather} \label{eq:likelihood}
    \Lagr (y_{obs}| \theta, M_i) = {\displaystyle \prod^{N_{obs}}_{k=1}} \dfrac{1}{\sqrt{2\pi \sigma_k^2}} \textrm{exp} \Big(- \dfrac{[y_{obs,k} - y_{mod,k}(\theta)]^2}{2 \sigma_k^2}\Big)
\end{gather}
where $N_{obs}$ is the number of observed data points (i.e., number of wavelength channels/data points in the observed spectrum), $\sigma_k$ is the standard deviation on the $k$-th data point (in our case its $(I/F)_{rms \ noise}$ as defined in eq. \ref{eq:rms_noise} and $y_{mod,k}(\theta)$ is the $k$-th simulated model data point (i.e., the simulated Hapke model spectrum binned to JIRAM's wavelength channels). \\

\noindent \emph{Prior distributions} \\

\noindent The prior function $p(\theta|M)$ or $\pi (\theta|M)$ in eq. \ref{eq:bayes_theorem} incorporates the prior knowledge we have of the parameters of our model. It also allows us to set bounds on the parameter space that will be explored during the posterior sampling. It is usually a good practice to assume a uniform distribution as the prior for the parameters, as they are relatively uninformative (as compared to, e.g., a Gaussian prior), which lets the data drive the solution. For the component abundances, a requirement is that they must sum up to unity. In a two component model, this is easy to enforce by having the abundance (in our case, the number density fraction) of one component, say $f_1$ as the free parameter that is uniformly distributed in the interval $[0,1]$

\begin{gather}
    p(f_1)= U(f_1) =  
\begin{dcases}
    1,& \text{for} f_1 \in [0,1]\\
    0,              & \text{otherwise}
\end{dcases}
\end{gather}
The abundance of the second component, say $f_2$, will then simply be equal to $1 - f_1$. For more than two components, the $\sum f_i = 1$ constraint can be satisfied using a Dirichlet distribution (eg. \cite{lapotre_compositional_2017}), but in this work we are only focusing on two components: amorphous and crystalline water-ice. 

To be agnostic about the distribution of grain sizes on Europa's surface, and to account for the multiple orders of magnitude that the grain sizes can span on Europa (eg. \citet{filacchione_serendipitous_2019} find grain sizes from tens to hundreds of microns), we use a log-uniform probability distribution

\begin{gather}
    p(\textrm{log}(D_i))=  
\begin{dcases}
    {\frac{1}{\textrm{log}(D_{max}) - \textrm{log}(D_{min})}},& {\text{for} \  \textrm{log}(D_i) \in [\textrm{log}(D_{max}), \textrm{log}(D_{min})]}\\
    { 0},              & \text{otherwise}
\end{dcases}
\end{gather}
where we set $D_{min}$ to be 10 microns and $D_{max}$ to be $10^3$ microns in all fitting exercises (sections \ref{sec:synthetic_and_lab} and \ref{sec:jiram}). These limits are motivated by previous works on reflectance data of Europa \citep[e.g.][]{cassidy_magnetospheric_2013,shirley_europas_2016, filacchione_serendipitous_2019}, where grains of 10s to 100s of microns have been reported. The lower limit of 10 microns is also motivated by the fact that our forward model operates in the geometric optics regime and hence the grain sizes need to be greater than the IR wavelength regime (2-5 $\mu$m) of our data. The upper limit of 10$^3$ microns is quite generous as the goal is to provide our framework a large parameter space to explore without imposing strict a priori expectations on possible solutions.

Finally, we assume a uniform distribution for the filling factor $\phi$

\begin{gather}
    p(\phi)=  
\begin{dcases}
    \dfrac{1}{\phi_{max} - \phi_{min}},& \text{for} \ \phi \in [\phi_{min}, \phi_{max}]\\
    0,              & \text{otherwise}
\end{dcases}
\end{gather}
where we set $\phi_{min} = 0.01$ and $\phi_{max} = 0.52$. The lower limit of 0.01 makes the porosity coefficient $K \approx 1$ (eq. \ref{eq:K_defn}), which reduces the radiative transfer equation (eq. \ref{eq:hapke_RT}) to its more familiar form without the parameter $K$ \citep[eg.][]{carlson_distribution_2005, clark_surface_2012}. The upper limit comes from \citet{hapke_theory_2012} where it is described as a critical value above which coherent effects become important and diffraction can't be ignored. \\

\noindent \emph{Exploring the parameter space: Nested sampling} \\

\noindent A probability distribution, like the posterior function in eq. \ref{eq:bayes_theorem}, can be approximated by using sampling algorithms like the popular Markov chain Monte Carlo or MCMC algorithm \citep{metropolis_equation_1953}. The posterior distribution of an individual parameter can then be obtained by marginalizing the approximate distribution over the other parameters. While MCMC has been extensively used in the Bayesian framework for planetary and exoplanetary spectroscopic retrieval analysis (e.g. \citet{lapotre_compositional_2017, rampe_sand_2018,irwin_nemesis_2008,madhusudhan_high_2011}), its Python implementations do not provide a computationally efficient way of evaluating the Bayesian evidence $\Z$. The Bayesian evidence is the multidimensional integral in eq. \ref{eq:bayes_theorem} and is directly used for Bayesian model comparison, discussed in the next section. Hence, we employ a different sampling technique called nested sampling \citep{skilling_nested_2006}, which directly computes the Bayesian evidence while generating posterior samples as a byproduct. We use the Python package \texttt{dynesty}\footnote[1]{dynesty.readthedocs.io}, which implements dynamic nested sampling \citep{higson_dynamic_2019}, a more computationally accurate version of the standard nested sampling algorithm. \\

\noindent \emph{Comparing models using Bayesian Model Comparison} \\

\noindent A key goal of our work is to compare how different models fit the data, instead of just assuming a ``correct" model from the start and reporting its parameter estimation results. Bayesian Model Comparison or BMC is a Bayesian alternative to the classical hypothesis testing that uses \textit{Bayes factors} to quantify the preference of one model over the other \citep{goodman_toward_1999}. It has been widely used in astronomy where low SNR data and little prior knowledge require sophisticated statistical tools (e.g. \citet{trotta_bayes_2008,yoon_new_2011,benneke_how_2013, macdonald_hd_2017}).
If we have to choose between two models $M_1$ and $M_2$ in light of our data $y_{obs}$, we can calculate the ratio between the probabilities of the two models as follows

\begin{gather} \label{eq:bayes_factor}
    \dfrac{p(M_1|y_{obs})}{p(M_2|y_{obs})} = \dfrac{\Z(y_{obs}|M_1)}{\Z(y_{obs}|M_2)} \dfrac{p(M_1)}{p(M_2)} = \B_{12} \dfrac{p(M_1)}{p(M_2)} = \B_{12}  \
\end{gather}
Here $\B_{12}$ is the \textit{Bayes factor} of Model 1 v/s Model 2, defined as the ratio of their Bayesian evidences $\Z$ (eq \ref{eq:bayes_theorem}). We also assume that the prior probability of both models is the same ($p(M_1) = p(M_2)$), as we have no reason to favour one model over the other.  A value of Bayes factor $\B_{12} > 1$ means that $M_1$ is more strongly favoured by the data rather than $M_2$. Generally speaking, Bayes factors $\B_{12}$ in the range of 1-2.5, 2.5-12, 12-150 and >150 would be considered as `marginal',`weak', `moderate' and `strong' evidence for model \citep{trotta_bayes_2008}, as shown in Table \ref{Tab: bmc}. The biggest advantage of using Bayesian model comparison is that it penalizes complex models and hence guards against overfitting. As explained in \citet{macdonald_hd_2017}, Bayesian model comparison favours models with high likelihood in a compact parameter space. Hence, the ability of Bayesian model comparison to take this into account is why it can be thought of as a generalization of Occam's Razor. 

The Bayes factor can also be related to the commonly used frequentist measure of $\sigma$-significance, using

\begin{gather} \label{eq:B_upper}
    \B_{12} \leq -\dfrac{1}{ep\textrm{ln}p}
\end{gather}

\begin{gather} \label{eq:p_to_sigma}
    p = 1 - \textrm{erf}\Big(\dfrac{N_\sigma}{\sqrt{2}}\Big)
\end{gather}
where $p$ is the `p-value', erf is the error function, and $N_\sigma$ is the `detection significance' \citep{sellke_calibration_2001, trotta_bayes_2008, benneke_how_2013}. Hence eq. \ref{eq:B_upper} converts the Bayes factor to an upper bound on p-value, which in turn gives a lower bound on the detection significance through eq. \ref{eq:p_to_sigma}. It should be noted that eq. \ref{eq:B_upper} is valid only for $p \leq e^{-1}$ or equivalently $\B_{12} \geq 1$. Table \ref{Tab: bmc} lists a range of Bayes factor values and their corresponding $\sigma$-significance values.

\begin{figure}[pos=htbp!]
    \centering
    \includegraphics[width=0.5\linewidth]{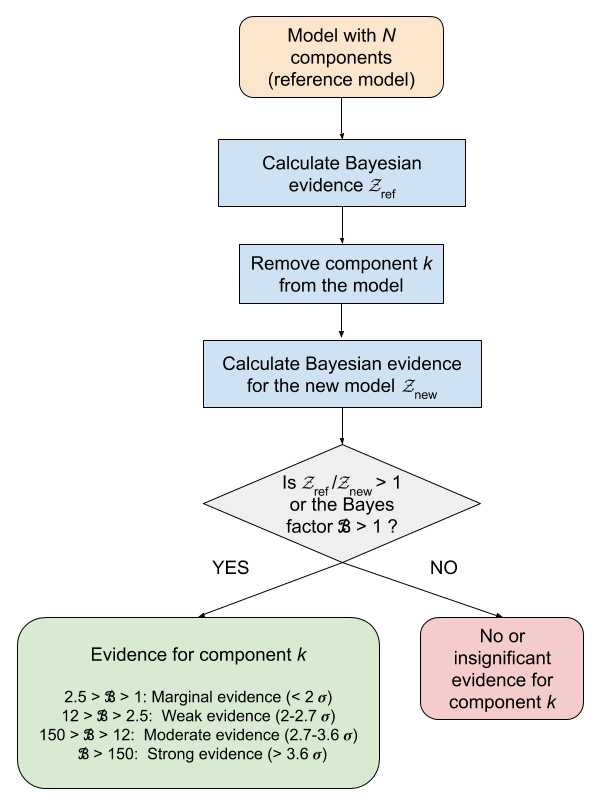}
    \caption{Process of evaluating the evidence of a particular species in our mixture, through Bayesian model comparison.}
    \label{fig:BMC}
\end{figure}

In planetary spectroscopic analysis, where we are trying to infer the composition of a surface (or an atmosphere) from its observed spectrum, one way of increasing the complexity of a model would be to add more components to the mixture. The Bayesian model comparison formulation we have described so far comes in very handy to quantify our confidence in having detected a particular constituent. For example, let's say we wish to quantify the presence of a constituent species $m$ in the observed spectrum. We can evaluate the Bayes factor of a more complex model that includes the species $m$ v/s a simpler model that does not have the species $m$. If the Bayes factor is strongly in favour of the more complex model, we can safely conclude that the species $m$ is present in the data. We can also quantify the degree of preference of the model with the species $m$ over the model without species $m$ through the $\sigma$-significance metric described in eqs. \ref{eq:B_upper}-\ref{eq:p_to_sigma}. In other words, we can evaluate the detection significance of species $m$ in the mixture. This process of evaluating the evidence for a particular species in the mixture is laid out in a flowchart in Figure~\ref{fig:BMC}. 

\begin{table}
\centering
\caption{Table \ref{Tab: bmc}: Translation Table between Frequentist Significance Values (p-values) and the Bayes Factor ($B_{12}$) when comparing two models $M_1$ and $M_2$. Adapted from \citet{benneke_how_2013}.}
\begin{tabular}{ |c c c c c| } 
  \hline
  $\B_{12}$ & ln$B_{12}$ & $p$-value & $\sigma$-significance & Interpretation \\
  \hline
  2.5 & 0.9 & 0.05 & 2.0$\sigma$ & \\
  2.9 & 1.0 & 0.04 & 2.1$\sigma$ & "Weak" detection \\
  8.0 & 2.1 & 0.01 & 2.6$\sigma$ & \\
  12 & 2.5 & 0.006 & 2.7$\sigma$ & "Moderate" detection \\
  21 & 3.0 & 0.003 & 3.0$\sigma$ & \\
  53 & 4.0 & 0.001 & 3.3$\sigma$ & \\
  150 & 5.0 & 0.0003 & 3.6$\sigma$ & "Strong" detection \\
  43000 & 11 & 6 x $10^{-7}$ & 5.0$\sigma$ & \\
  \hline
\end{tabular}
\label{Tab: bmc}
\end{table}

\section{Application to synthetic and laboratory data} \label{sec:synthetic_and_lab}

\vspace{10pt}

\subsection{Tests with synthetic data} \label{sec:synthetic_data}

\vspace{10pt}

\begin{figure}[pos=htbp!]
    \centering
    \includegraphics[width=0.75\linewidth]{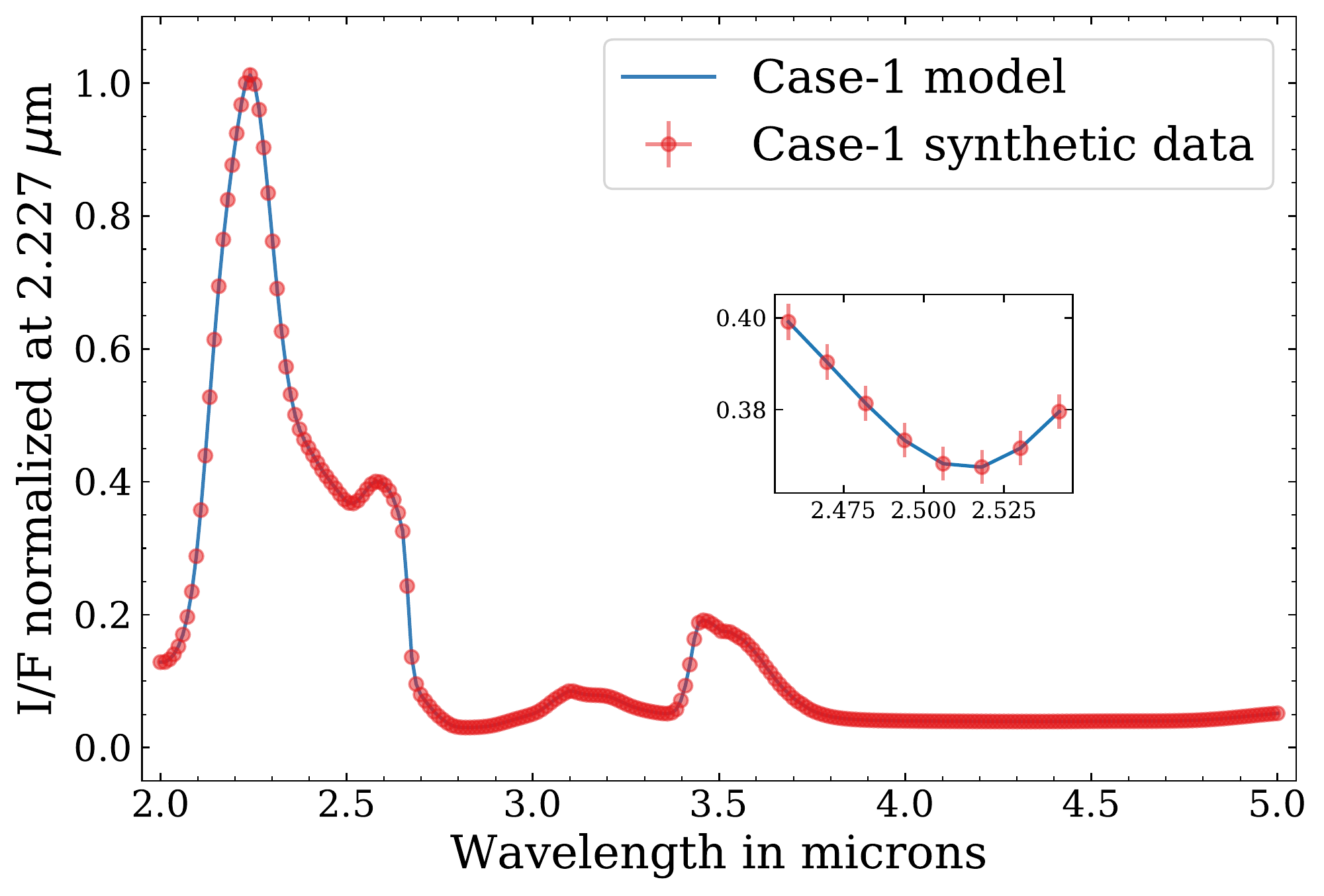}
    \caption{Synthetic spectrum (blue) and the corresponding binned synthetic data points (red), generated using a two-component intimate-mixture Hapke model of amorphous and crystalline ice with parameters: incidence angle $= 45.0$ deg., emission angle $= 45.0$ deg., phase angle $g = 90$ deg., grain size of amorphous ice = $D_{am} = 200 \ \mu$m, Grain size of crystalline ice = $D_{cr} = 100 \ \mu$m, number density fraction of amorphous ice $\textrm{f}_{am}= 0.5$, number density fraction of crystalline ice $\textrm{f}_{cr} = 1 - \textrm{f}_{am} = 0.5$. The error bars correspond to an SNR=100 and the inset plot zooms in near 2.5 $\mu$m to show the tiny error-bars of the order 0.001 in $I/F$ units.}
    \label{fig:syn_data_1}
\end{figure}

\begin{figure}[pos=htbp!]
    \centering
    \includegraphics[width=0.5\linewidth]{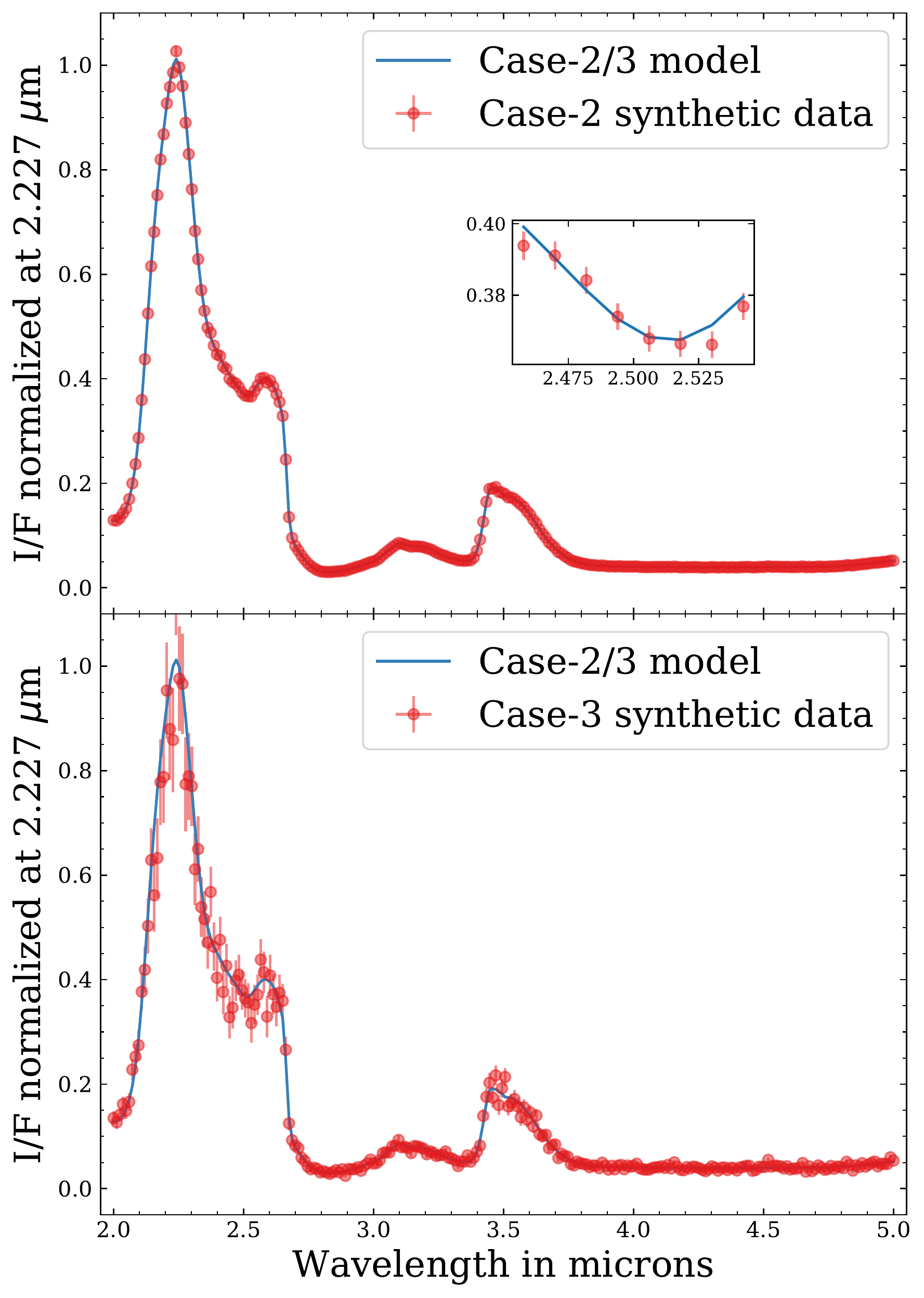}
    \caption{Synthetic data (red points) for Case-2 (top) and Case-3 (bottom), generated using a two-component intimate-mixture Hapke model of amorphous and crystalline ice with parameters same as those in Case-1 (Figure~ \ref{fig:syn_data_1}). In these cases however, a Gaussian noise/scatter has been added to the synthetic data points.  The inset plot in the top panel zooms in near 2.5 $\mu$m to show the scatter around the true model in blue. Case-2 corresponds to an SNR=100 while Case-3 corresponds to an SNR=10. The common synthetic spectrum used to generate these synthetic data is shown in blue in both plots.}
    \label{fig:syn_data_2}
\end{figure}

\noindent The simplest way to check if a fitting-analysis framework is working correctly is to evaluate its performance on synthetic data. A synthetic dataset in this case would start with a spectrum generated using the forward model of the analysis framework, which in our case is the Hapke reflectance model. We can then define an SNR and add noise or scatter to the synthetic data points. A retrieval analysis on this synthetic data should be able to accurately retrieve the true values of the free parameters in our analysis. We perform this test using three different synthetic spectra:

\begin{enumerate}[label={Case-\arabic*}]
    \item: A two-component intimate-mixture model with amorphous and crystalline ice, with error-bars but without scatter. In other words, the synthetic data points lie perfectly centered on the model used to generate them. As described in \citet{feng_characterizing_2018}, the attraction of testing a retrieval framework on such data is that a retrieval without Gaussian scatter should have posterior parameter distributions centred on the true parameters. \citet{feng_characterizing_2018} showed that a posterior parameter distribution obtained in this case is the average of many independent retrievals for different noise instances (via the central limit theorem), and hence the posteriors here will not be biased by any one particular noise instance.
    
    Figure~ \ref{fig:syn_data_1} shows this synthetic data, with the fixed parameters of the model listed in the Figure~ caption. The synthetic was also normalized at 2.227 $\mu$m, similar to the JIRAM data (see top panel of Figure~ \ref{fig:data}). We chose an SNR of 100 to best-replicate the SNR of the JIRAM data beyond 3 $\mu$m (see bottom panel of Figure~ \ref{fig:data}). JIRAM's SNR decreases at larger wavelengths due to intrinsic low reflectance of Europa and worse instrumental sensitivity \citep{filacchione_serendipitous_2019}. 
    \item: A two-component intimate-mixture model with amorphous and crystalline ice, generated using same parameters as Case-1 but now with Gaussian scatter or noise added, as shown in the top panel of Figure~ \ref{fig:syn_data_2}.
    \item: Same as Case-2 but with a lower SNR of 10, as shown in the bottom panel of Figure~ \ref{fig:syn_data_2}. This SNR is representative of the typical Galileo/NIMS observations of Europa, whose SNR ranges from 5-50 \citep{greeley_future_2009}. 
\end{enumerate}

For the primary retrieval analysis on these three synthetic datasets, we use the following free parameters, with uniform prior functions on intervals that insures their physical relevance (see section \ref{sec:bayesian_framework}):

\begin{enumerate}
    \item the common logarithm of amorphous ice grain size (in microns) or $\mathrm{log}_{10}(D_{am})$, with prior bounds of (1.0, 3.0)
    \item the common logarithm of crystalline ice grain size (in microns) or $\mathrm{log}_{10}(D_{cr})$, also with prior bounds of (1.0, 3.0)
    \item the grain number density fraction of amorphous ice $f_{am}$, with prior bounds of (0, 1)
    \item the filling factor $\phi$, with prior bounds of (0.01, 0.52)
\end{enumerate}

\begin{figure}[pos=htbp!]
    \centering
    \includegraphics[width=0.75\linewidth]{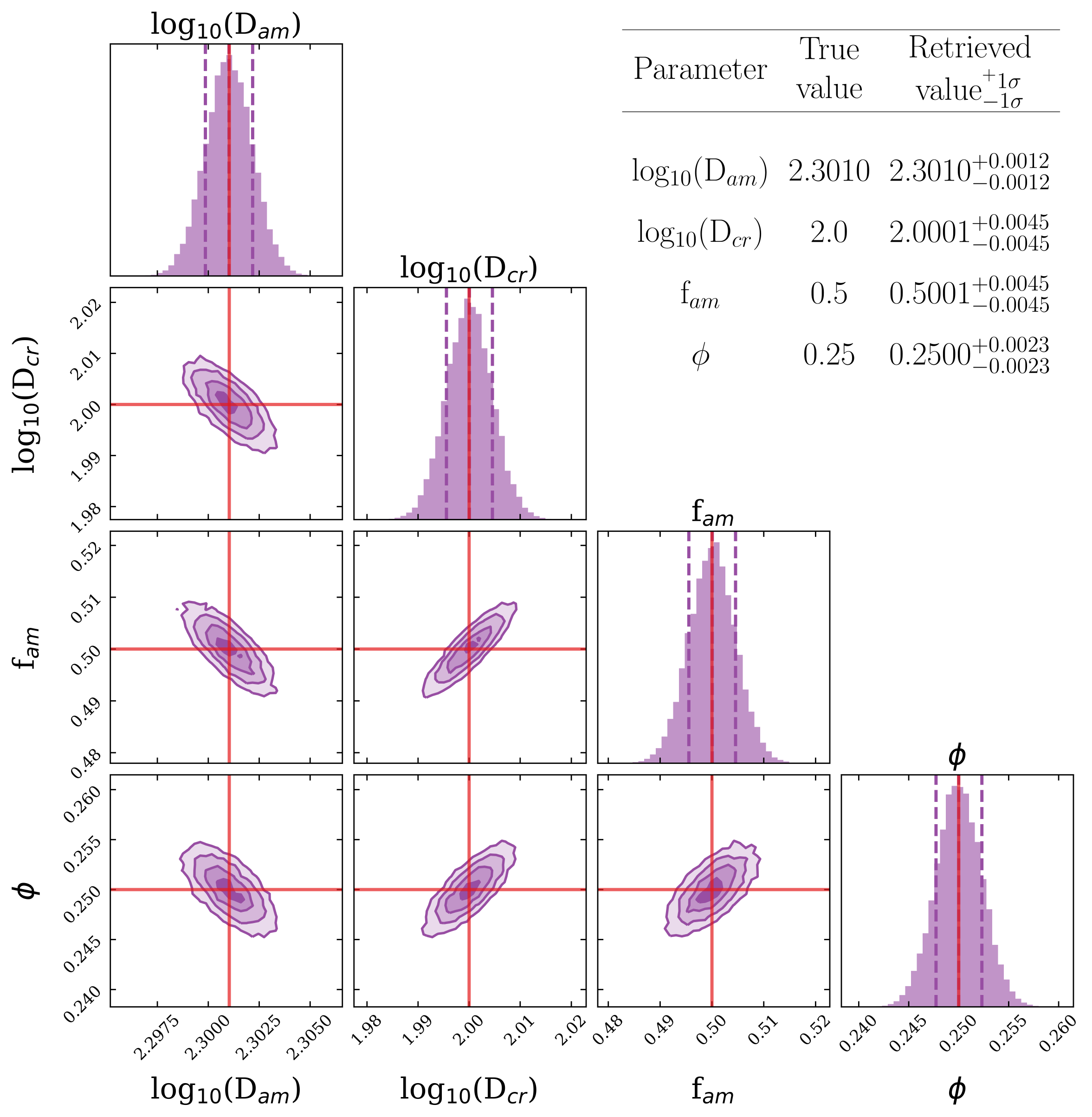}
    \caption{The `corner' plot shows samples generated from our Bayesian retrieval analysis of the synthetic data in Case-1 (Figure~ \ref{fig:syn_data_1}) with the two-component intimate mixing model. The histograms show the marginalized probability distributions of the common logarithm of amorphous ice grain size (in microns) or $\mathrm{log_{10}(D_{am})}$, the common logarithm of crystalline ice grain size (in microns) or $\mathrm{log_{10}(D_{cr})}$, the grain number density fraction of amorphous ice $f_{am}$ and the filling factor $\phi$ respectively, from top to bottom.  The red lines show the true parameter values while the dashed purple lines are the 1-$\sigma$ upper and lower limit (68 \% confidence interval). The inset table lists the true values and retrieved median values with the associated 1-$\sigma$ upper and lower limits for all four parameters. The other six plots are pair-wise 2D distributions, that illustrate the correlations between parameters. The contours in these 2D distributions correspond to 0.5, 1, 1.5, and 2-$\sigma$ intervals.}
    \label{fig:syn_post_1}
\end{figure}

\begin{figure}[pos=htbp!]
    \centering
    \includegraphics[width=\linewidth]{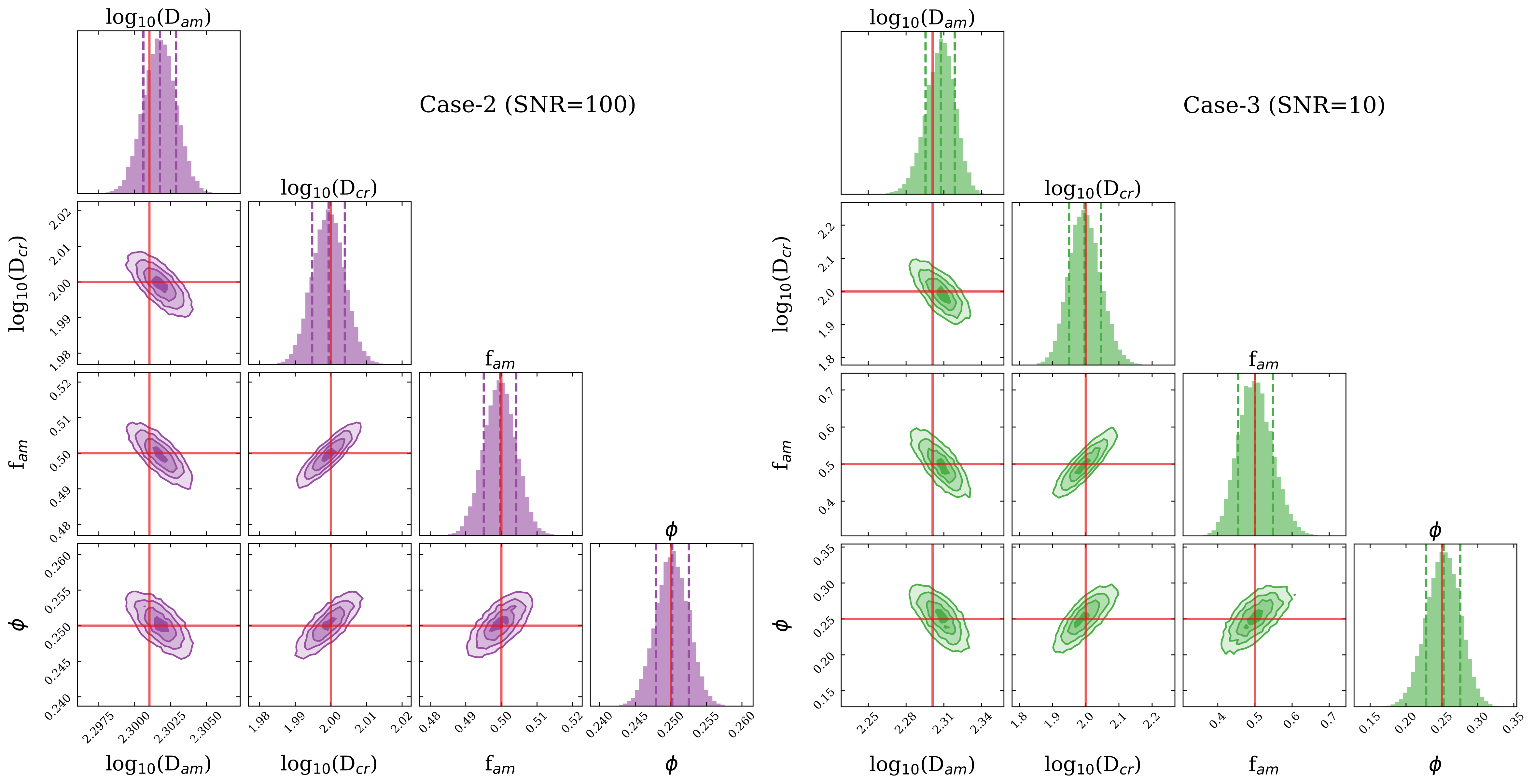}
    \caption{The plot shows samples generated from our Bayesian retrieval analysis of the synthetic data in Case-2 (left) and Case-3 (right) (both from Figure~ \ref{fig:syn_data_2}), with the two-component intimate mixing model. Despite reducing the SNR by a factor of ten from Case-2 to Case-3, the retrieved parameter distributions remain within 1-$\sigma$ of the true values. However, due to the much higher SNR the parameter distributions are much more sharply peaked for Case-2 as compared to Case-3. The median parameter values and their 1-$\sigma$ limits are listed in row 1 of Table \ref{Tab: syn_results}.} 
    \label{fig:syn_post_2}
\end{figure}

The resulting parameter posterior distributions of Case-1 and Cases 2 \& 3 are shown in the `corner' plots in Figures~\ref{fig:syn_post_1} and \ref{fig:syn_post_2} respectively. For Case-1, since there is no scatter in the synthetic data, the model that was used to generate the synthetic data fits the data perfectly, i.e., with a chi-squared value of 0. Hence, one should expect very sharp posteriors for all four parameters, with peaks very close to the true parameter values. This is indeed what we see in Figure~ \ref{fig:syn_post_1}. The derived posterior distributions for Case-2, which is noisier than Case-1, and Case-3, which is noisier and has a lower SNR than both Case-1 and Case-2, are also similarly sharply peaked and well-constrained around the true parameter values as shown in Figure~ \ref{fig:syn_post_2}. This shows the robustness of Bayesian inference against noise in the data.

\noindent Another interesting output in the three corner plots, Figure~s \ref{fig:syn_post_1} and \ref{fig:syn_post_2}, are the pairwise 2D distributions, which show the correlations between different parameters. For example, it's clear that the grain-size parameter for amorphous ice, $\mathrm{log}_{10}(D_{am})$, is anti-correlated with its abundance $f_{am}$. This can be qualitatively explained as follows: in order to maintain the net absorption from amorphous ice, increasing the grain size of particles would require us to decrease their abundance, as larger grains absorb more light. The positive correlation between the grain-size parameter of crystalline ice, $\mathrm{log}_{10}(D_{cr})$, and $f_{am}$ follows from this. Increasing $f_{am}$ would mean decreasing the abundance of crystalline ice, which means to achieve the same net absorption by crystalline ice we would need to make its grains larger in size. The correlation of filling factor $\phi$ with the other three parameters also intuitively stems from its effect on the amplitude of the spectrum. For example, it is positively correlated with $f_{am}$ because while increasing $f_{am}$ for the larger amorphous ice grains causes more absorption and decreases overall reflectance of the mixture, increasing $\phi$ increases reflectance, as seen in Figure~ \ref{fig:sensitivity}.

For the data in Case-2 and Case-3, we also performed retrieval analyses with three other models: two-component linear mixing model (discussed in section \ref{sec:forward_model}), amorphous-ice-only model and crystalline-ice-only model. The results of these analyses, along with those of the intimate-mixing model, are shown in Table \ref{Tab: syn_results}. For Case-2, where the data has an SNR of 100, the Bayes factor of the intimate mixing model v/s other three models is very high, which indicates that the intimate mixing model is strongly preferred by the Bayesian framework. We can quantify this preference by using the scheme of evaluating $\sigma-$significances as described in eq. \ref{eq:B_upper} and eq. \ref{eq:p_to_sigma}. We find that the intimate-mixing model is preferred over the linear-mixing model at a strong confidence of 19.94$\sigma$ (see Table \ref{Tab: bmc}). The Bayes factor values for the intimate mixing model v/s the single component models can be employed to find the detection significance of amorphous and crystalline ice individually, as described earlier and in the flowchart in Figure~ \ref{Tab: bmc}. Hence, the amorphous-ice only model, that is obtained from removing crystalline ice from the two-component intimate-mixing model, tells us the detection significance or confidence of crystalline ice. Similarly, the Bayes factor for the crystalline-ice only model can give us the detection significance of amorphous ice in the mixture. We see that for both these cases, the $\sigma$-significances are $> 30\sigma$, indicating very high confidence for the presence of both forms of ices in the mixture.

The parameter estimations for the lower SNR Case-3 data analysis with the intimate-mixing model also agree with the true values, within their $1-\sigma$ limit. We see that the intimate-mixing model is once again preferred over the other three models, albeit with a lesser degree of preference or $\sigma$-significance as compared to the Case-2 results. Finally, also shown are the reduced chi-squared, $\chi^2_{reduced}$, values of the best-fitting spectra of the four models. For Case-2, this metric is closest to 1 for the intimate-mixing model, indicating that the best-fitting intimate-mixing model has a good fit to the data. For Case-3 however, since it is noisier,  it is interesting to note that the  $\chi^2_{r,bf}$ values are basically the same for the intimate-mixing and linear-mixing models' best-fit solutions. However, the Bayes factor shows us that intimate mixing is preferred, albeit at a `weak' confidence of $2.23\sigma$. Hence, comparing just the $\chi^2_{r,bf}$ of the best-fit solutions of the two models would have been misleading. The Bayesian evidence, since it is integrated over the entire parameter space, is a more comprehensive metric by which to compare models.

\begin{table}
\centering
\captionsetup{width=0.9\textwidth}
\caption{Table \ref{Tab: syn_results}: Bayesian inference results for fitting synthetic data of Case-2 and Case-3 are presented in Figure~ \ref{fig:syn_data_2}, corresponding to SNR = 100 and 10 respectively. The true parameters used to generate the synthetic data were log$_{10} D_{am}$=2.3010, log$_{10} D_{cr}$=2.0, $f_{am}$=0.5 and $\phi$=0.25. From left to right, the columns specify the model, free parameters, bounds on the prior, the retrieved median parameter values, the natural logarithm of Bayes factor with respect to the intimate-mixing model, ln$(\boldsymbol{\B_{0,i}})$ (eq. \ref{eq:bayes_factor}), and the reduced chi-squared value of the best-fitting model $\boldsymbol{\chi^2_{r, bf}}$. Next to ln$(\boldsymbol{\B_{0,i}})$, in brackets, the $\sigma$-significance for the preference of the intimate-mixing model (which is the reference model) over the other three models is mentioned. `> 30$\sigma$' preference for some cases stems from ln$(\boldsymbol{\B_{0,i}})$ being very high.}
 \resizebox{0.9\textwidth}{!}{\begin{minipage}{\textwidth}
 \hspace*{-1.25cm}\begin{tabular}{p{2.5cm}  p{1.5cm}  p{2cm} p{2.5cm} p{1.5cm} p{1cm} p{2.5cm} p{1.5cm} p{1cm}}
  \hline
  \textbf{Model} & \makecell{\textbf{Free} \\ \textbf{params}} & \makecell{\textbf{Prior} \\ \textbf{bounds}} & \makecell{\textbf{Case 2:} \\ \textbf{Solution}}  & \makecell{\textbf{ Case-2:} \\ ln$(\boldsymbol{\B_{0,i}})$} & \textbf{Case-2: }$\boldsymbol{\chi^2_{r, bf}}$ & \makecell{\textbf{Case-3:} \\ \textbf{Solution}}  & \makecell{\textbf{Case 3:} \\ ln$(\boldsymbol{\B_{0,i}})$} & \textbf{Case-3: }$\boldsymbol{\chi^2_{r, bf}}$\\
  \hline
 \makecell{Am. \& Cr. ice \\ (intimate mixing)}  & \makecell{log$_{10} D_{am}$,  \\ log$_{10} D_{cr}$, \\ $f_{am}$, $\phi$}  & \makecell{(1.0,3.0), \\ (1.0,3.0), \\ (0,1), \\ (0.01,0.52)} & \makecell{2.3017$_{-0.0011}^{+0.0012}$ \vspace{5pt}\\ 1.9994$_{-0.0045}^{+0.0044}$ \vspace{5pt} \\ 0.4996$_{-0.0045}^{+0.0047}$ \vspace{5pt} \\ 0.2502$_{-0.0022}^{+0.0023}$ \vspace{5pt}} &  ref. & 0.94 & \makecell{2.3076$_{-0.0122}^{+0.0110}$ \vspace{5pt} \\    1.9963$_{-0.0465}^{+0.0500}$ \vspace{5pt} \\ 0.5001$_{-0.0455}^{+0.0485}$ \vspace{5pt}  \\ 0.2523$_{-0.0243}^{+0.0232}$ \vspace{5pt}} & ref. & 0.94 \\
  \hline
 \makecell{Am. \& Cr. ice \\ (linear mixing)} & \makecell{log$_{10} D_{am}$,  \\ log$_{10} D_{cr}$, \\ $f_{am}$, $\phi$}  & \makecell{(1.0,3.0), \\ (1.0,3.0), \\ (0,1), \\ (0.01,0.52)} & \makecell{2.3149$_{-0.0011}^{+0.0011}$ \vspace{5pt} \\ 1.9462$_{-0.0063}^{+0.0063}$ \vspace{5pt} \\ 0.8623$_{-0.0022}^{+0.0022}$ \vspace{5pt} \\ 0.2328$_{-0.0026}^{+0.0025}$ \vspace{5pt}} & \makecell{195.79 \\ (19.94 $\sigma$)} & 2.53 & \makecell{2.3197$_{-0.0118}^{+0.0103}$ \vspace{5pt} \\ 1.9392$_{-0.0683}^{0.0689}$ \vspace{5pt} \\ 0.8670$_{-0.0237}^{+0.0205}$ \vspace{5pt} \\ 0.2352$_{-0.0273}^{+0.0258}$ \vspace{5pt}} & \makecell{1.36 \\ (2.23 $\sigma$)} & 0.95 \\
  \hline
  Am. ice only & \makecell{log$_{10} D_{am}$, \\ $\phi$}  & \makecell{(1.0,3.0), \\ (0.01,0.52)} & \makecell{2.3027$_{-0.0006}^{0.0007}$ \vspace{5pt} \\ 0.1846$_{-0.0015}^{+0.0014}$ \vspace{5pt}} & \makecell{12573.88 \\ (> 30$\sigma$)}  & 102 & \makecell{2.3074$_{-0.0064}^{+0.0064}$ \vspace{5pt} \\ 0.1902$_{-0.0156}^{+0.0154}$ \vspace{5pt}}  & \makecell{115.13 \\ (15.36 $\sigma$)} & 1.90 \\
  \hline
  Cr. ice only  &  \makecell{log$_{10} D_{cr}$, \\ $\phi$}  & \makecell{(1.0,3.0),\\ (0.01,0.52)} & \makecell{2.3587$_{-0.0008}^{+0.0007}$ \vspace{5pt}\\ 0.2954$_{-0.0016}^{+0.0015}$ \vspace{5pt}} &  \makecell{86145.68 \\ (> 30$\sigma$)} & 696 &  \makecell{2.3646$_{-0.075}^{+0.084}$ \vspace{5pt} \\ 0.2971$_{-0.0158}^{+0.0146}$ \vspace{5pt}} & \makecell{846.34 \\ (> 30$\sigma$)} & 7.80 \\
  \hline
\end{tabular}
\label{Tab: syn_results}
\end{minipage}}
\end{table}

\subsection{Tests with laboratory data} \label{sec:lab}

\vspace{10pt}

\begin{figure}[pos=htbp!]
    \centering
    \includegraphics[width=0.5\linewidth]{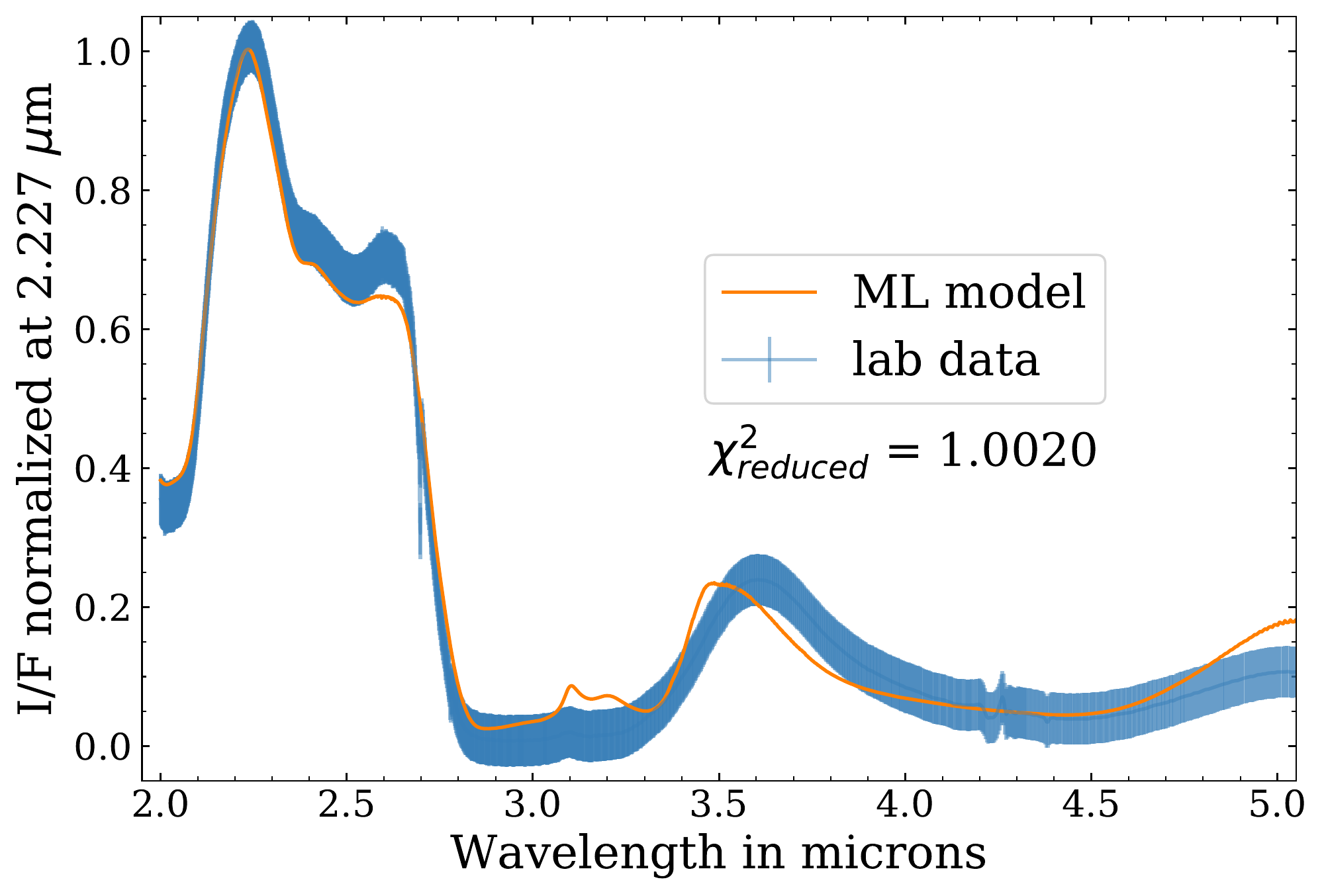}
    \caption{The normalized (at 2.227 $\mu$m) \citet{clark_surface_2012} laboratory data (blue), with the median retrieved error-bars (see parameter distributions in Figure~ \ref{fig:lab_post}), and the normalized (at 2.227$\mu$m) maximum-likelihood model (orange). The reduced chi-square ($\chi^2_{reduced}$) value of this fit is 1.002, indicating an excellent agreement between the model and the data.}
    \label{fig:lab_data_ML}
\end{figure}

We obtained a laboratory reflectance spectrum of crystalline ice grains sample at 123 K, as used in Fig 20. of \citet{clark_surface_2012}, (R. Clark, personal communication). This data, normalized with its value at $2.227 \ \mu$m, is shown in Figure~\ref{fig:lab_data_ML}. We have normalized the spectrum, like we did with synthetic data in the previous section, since we wish to test our analysis framework's ability to retrieve parameters from normalized data. The sample consists of grains of around 20 microns (the exact distribution of grain-sizes was not documented). It was observed such that the light source, which subtends a beam of around 8$\degree$, was shone at the sample with an incidence angle of 20$\degree$ and phase angle of 20$\degree$. The emission angle of the observation was $35\degree$. For a Bayesian analysis of this data with a crystalline-ice-only Hapke model, we let the following parameters be free:

\begin{itemize}
    \item the common logarithm of crystalline ice grain size (in microns) or $\mathrm{log}_{10}(D_{cr})$, also with prior bounds of (0.5, 3.0)
    \item the filling factor $\phi$, with prior bounds of (0.01, 0.52)
    \item the natural logarithm of the error/uncertainty $\mathrm{log}_{e}\sigma$, common for all data points, with prior bounds of (-6.0, -1.0). 
\end{itemize}

Here we treat the errors/uncertainties on the lab spectrum as a free parameter for our fitting exercise, assuming that all the data points in the spectrum have the same error, i.e., the errors are assumed to be the same across all wavelength channels. This agrees with the general characteristics of the instrument, whose noise levels don't vary much in the 2-5 $\mu$m wavelength range. This error parameter not only accounts for the uncertainties in the data due to instrumental noise, but also the limitations/missing physics of our forward model \citep{hogg_data_2010, line_uniform_2015}. This approach also allows the uncertainties in the error parameter to be properly marginalized into the probability distributions of the other parameters. The error parameter $\sigma$ appears in two places in the likelihood function in eq. \ref{eq:likelihood}: 1) The term inside the exponential 2) the Gaussian normalization factor, $1/\sqrt{2 \pi \sigma^2}$. The term inside the exponential is the familiar `chi-square' that penalizes large residuals. The Gaussian normalization factor provides a balance to the chi-square term and prevents it from approaching infinity if we reduce the errors $\sigma$.

\begin{figure}[pos=htbp!]
    \centering
    \includegraphics[width=0.75\linewidth]{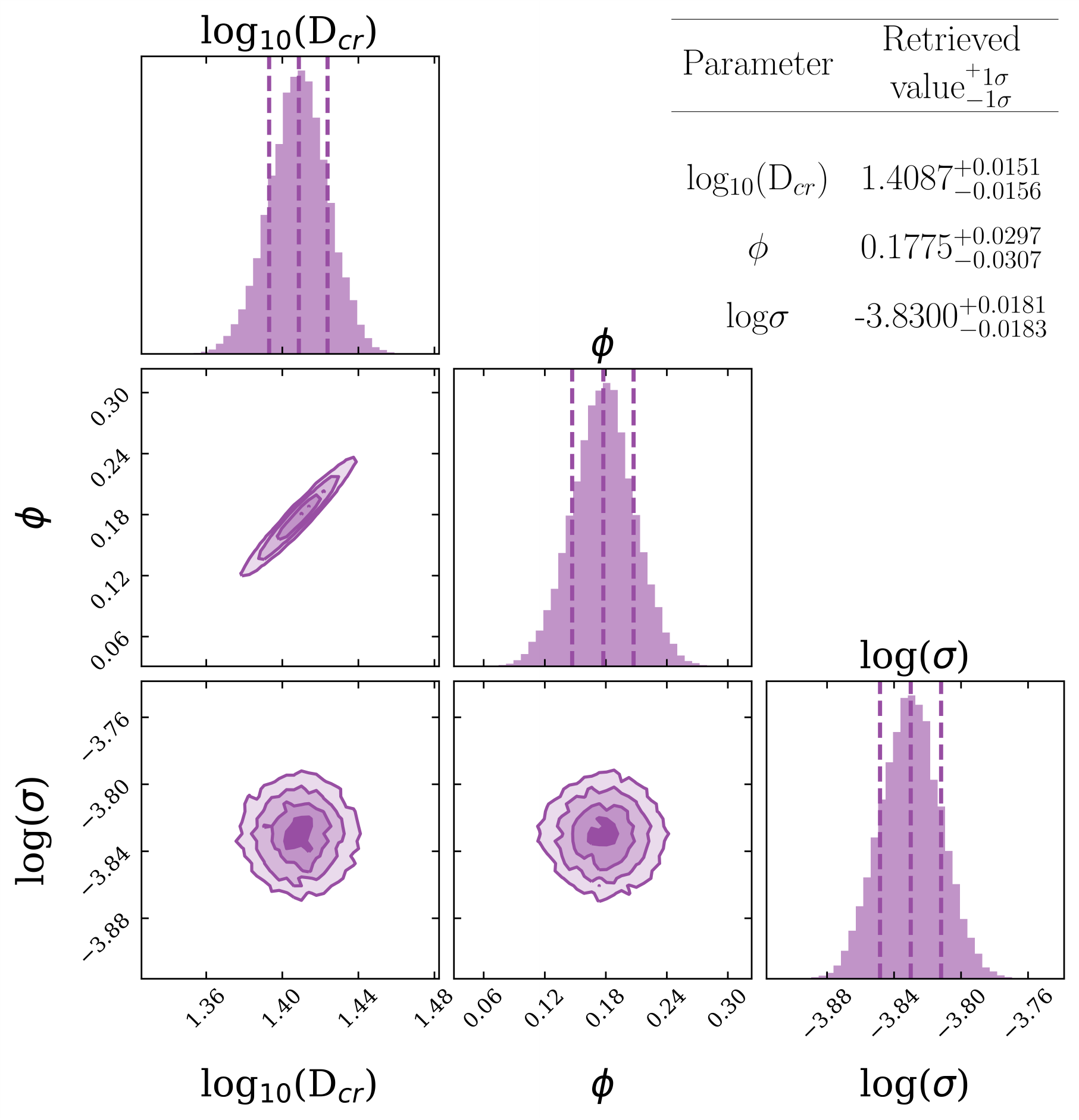}
    \caption{Parameter distributions and correlations for the Bayesian analysis of the lab data, shown in Figure~ \ref{fig:lab_data_ML}, with the one component Hapke model with crystalline ice. The distribution of grain sizes we derive peaks at around 29 microns, which is in good agreement with the estimated true value of around 20 microns of the grains in the lab samples.}
    \label{fig:lab_post}
\end{figure}

The parameter distributions for the Bayesian analysis of these lab data is shown in Figure~ \ref{fig:lab_post}. All three parameters in our fitting analysis are well constrained. The median of the grain size parameter $\mathrm{log}_{10}(D_{cr})$ distribution is 1.4087, which is equivalent to $D_{cr}=$25.62 microns. This is remarkably close to the actual (rough) value of around 20 microns for the grains in the lab sample, despite the simplifying assumption of a single, average grain-size (instead of a distribution of grain sizes) that is used in Hapke's model. The error parameter distribution $\mathrm{log}_{e}\sigma$ has a median value of -3.830, which is equivalent to $\sigma=0.021$. One way to check if this estimate is correct is to inspect the residuals of the maximum-likelihood model's fit to the data (Figure~ \ref{fig:lab_data_ML}). The standard deviation of the residuals of a good fitting model to any given data should be close in value to the error bars on the data. In our case, the standard deviation of the residuals in Figure~ \ref{fig:lab_data_ML} is equal to the error estimated by the Bayesian inference to 4 significant digits (=0.021). The filling factor $\phi$ has been constrained at 0.1775$_{-0.0307}^{+0.0297}$, which is indicative of a porous material. This is what we expect as the lab sample consisted of loose grains. Hence $\phi$, which is also included in the analysis of JIRAM data in the next section, has been validated to work correctly. To further check the robustness of these results, we set the error to be constant, equal to 0.021, and re-run the retrieval analysis with $\mathrm{log}_{e}\sigma$ and $\phi$ as the only free parameters. The new parameter distributions for $\mathrm{log}_{e}\sigma$ and $\phi$ were very similar to the distributions is Figure~ \ref{fig:lab_post}, with the median values being equal to 4 significant digits (not shown). Figure~ \ref{fig:lab_data_ML} also shows that the maximum-likelihood model, generated from the highest probability parameters from Figure~ \ref{fig:lab_post}, does a very good job fitting the data ($\chi_{reduced}^2 = 1.0029$). There are minor discrepancies around 2.2, 3.0 and 3.5 $\mu$m, which may result from limitations in the simple analytical treatment of reflection in the Hapke model \citep{clark_surface_2012}. 


\section{Application to JIRAM data} \label{sec:jiram}

Our framework has been shown to perform robustly on a range of synthetic and laboratory data. We now move on with confidence in our approach to analyze the \textit{Juno}/JIRAM data of Europa's surface (Figure~ \ref{fig:data}). Similar to section \ref{sec:synthetic_and_lab}, we set-up the analysis on the JIRAM  data sets with a two-component model that has the following free parameters: 

\begin{enumerate}
    \item the common logarithm of amorphous ice grain size (in microns) or $\mathrm{log}_{10}(D_{am})$, with prior bounds of (1.0, 3.0)
    \item the common logarithm of crystalline ice grain size (in microns) or $\mathrm{log}_{10}(D_{cr})$, also with prior bounds of (1.0, 3.0)
    \item the grain number density fraction of amorphous ice $f_{am}$, with prior bounds of (0, 1)
    \item the filling factor $\phi$, with prior bounds of (0.01, 0.52)
\end{enumerate}

We are using the optical constants from \citet{mastrapa_optical_2009} at 120 K, which is the commonly used temperature in the literature for spectroscopic analysis of Europa (e.g. \citet{carlson_distribution_2005}). Although water-ice spectral features are dependent on temperature \citep{mastrapa_optical_2009}, especially the 3.6 $\mu$m feature \citep{filacchione_saturns_2016}, the JIRAM data in this region has a lot of scatter and is mostly flat (see Figure \ref{fig:data}). In fact, as will be discussed later, the mismatch between our best-fit model and the data in the 3.6 $\mu$m region is large enough that the position of the peak should not matter (\ref{fig:jiram_im_lm_ML}). Hence, the position of this peak in our spectral models, which arises from the choice of temperature of the water-ice optical constants (between 80-130 K \citep{spencer_temperatures_1999, filacchione_serendipitous_2019}) should not affect our results. Next, the observation geometry parameters for the model are as stated in Table \ref{Tab: data}, with incidence angle equal to $28.5\degree$, emission angle equal to $73.4\degree$ and phase angle equal to $91.5\degree$. Also, we set the internal scattering coefficient $s=0$, for reasons detailed in section \ref{sec:bayesian_framework}.

Similar to the simulated data example of section \ref{sec:synthetic_and_lab} (see Table \ref{Tab: syn_results}), we perform our analysis for the two different mixing schemes, \textit{intimate mixing} (IM) and \textit{linear mixing} (LM). Both mixing schemes have been used in previous studies using the Galileo/NIMS of Europa, and in a review study \citet{shirley_europas_2016} found that they return similar abundances. It should be noted that the locations analyzed by \citet{shirley_europas_2016} are beyond 90$\degree$W longitude, while the data we are analyzing come from a different region, around 40$\degree$W longitude (see Table \ref{Tab: data}).

\begin{figure}[pos=htbp!]
    \centering
    \includegraphics[width=0.75\linewidth]{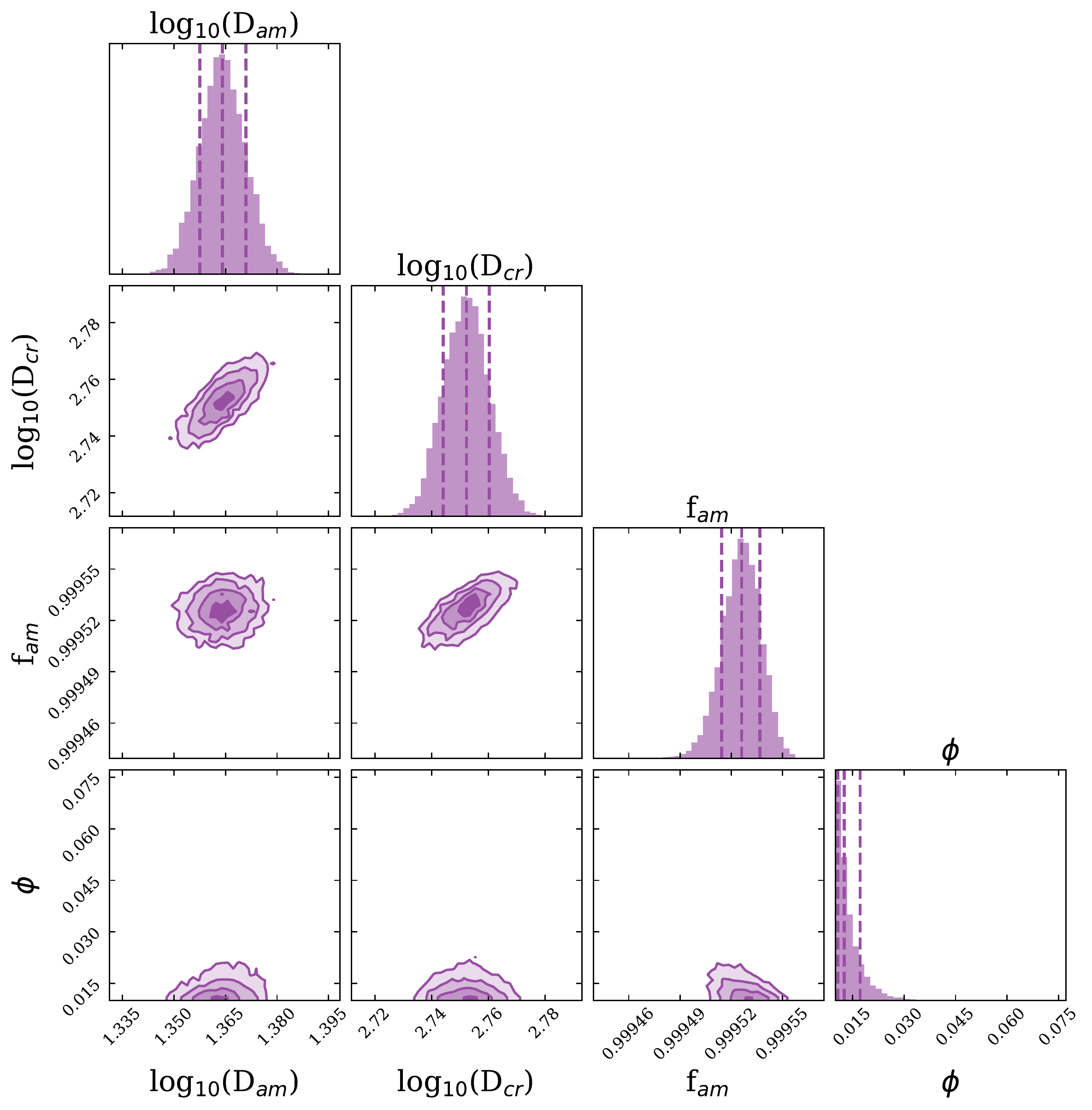}
    \caption{Parameter distributions and correlations for the two-component amorphous plus crystalline ice model retrieval on JIRAM data, with intimate mixing scheme (eq. \ref{eq:im_eqn}). The notations are the same as used in Figure~s \ref{fig:syn_post_1} and \ref{fig:syn_post_2}. The prior function for all parameters was uniform over the range (1.,3.) for log$_{10}D_{am}$, (1.,3.) for log$_{10}D_{cr}$, (0,1) for $f_{am}$ and (0.01,0.52) for $\phi$. The retrieved parameter values are shown in the first row of Table \ref{Tab: jiram_results}.}
    \label{fig:jiram_im}
\end{figure}

\begin{figure}[pos=htbp!]
    \centering
    \includegraphics[width=0.75\linewidth]{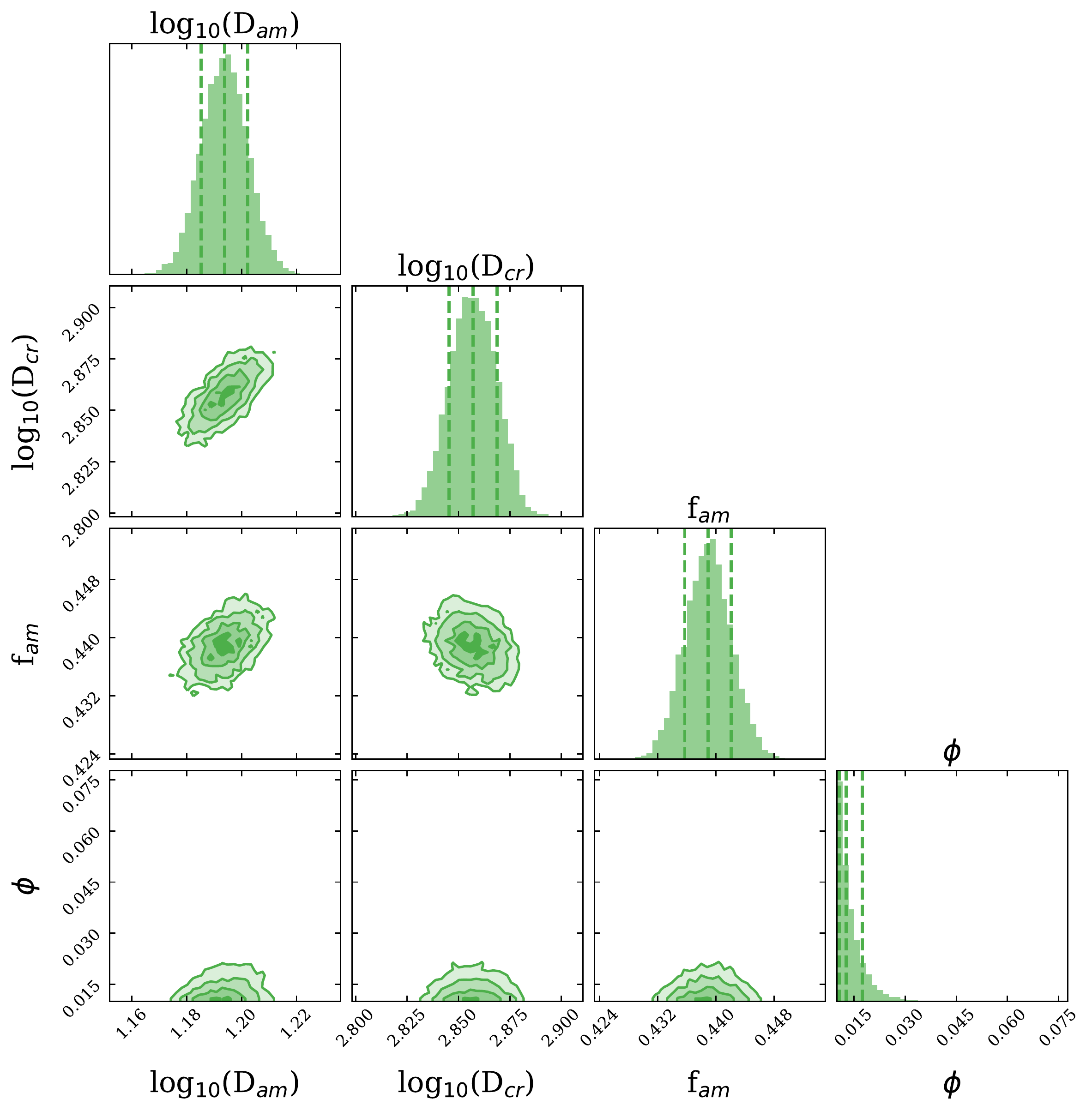}
    \caption{Parameter distributions and correlations for the two-component amorphous plus crystalline ice model retrieval on JIRAM data, with linear mixing scheme (eq. \ref{eq:lm_eqn}). The notations are the same as used in Figure~s \ref{fig:syn_post_1}, \ref{fig:syn_post_2} and \ref{fig:jiram_im}. The prior function for all parameters was uniform over the range (1.,3.) for log$_{10}D_{am}$, (1.,3.) for log$_{10}D_{cr}$, (0,1) for $f_{am}$ and (0.01,0.52) for $\phi$. The retrieved parameter values are shown in the second row of Table \ref{Tab: jiram_results}.}
    \label{fig:jiram_lm}
\end{figure}

\begin{figure}[pos=htbp!]
    \centering
    \includegraphics[width=0.5\linewidth]{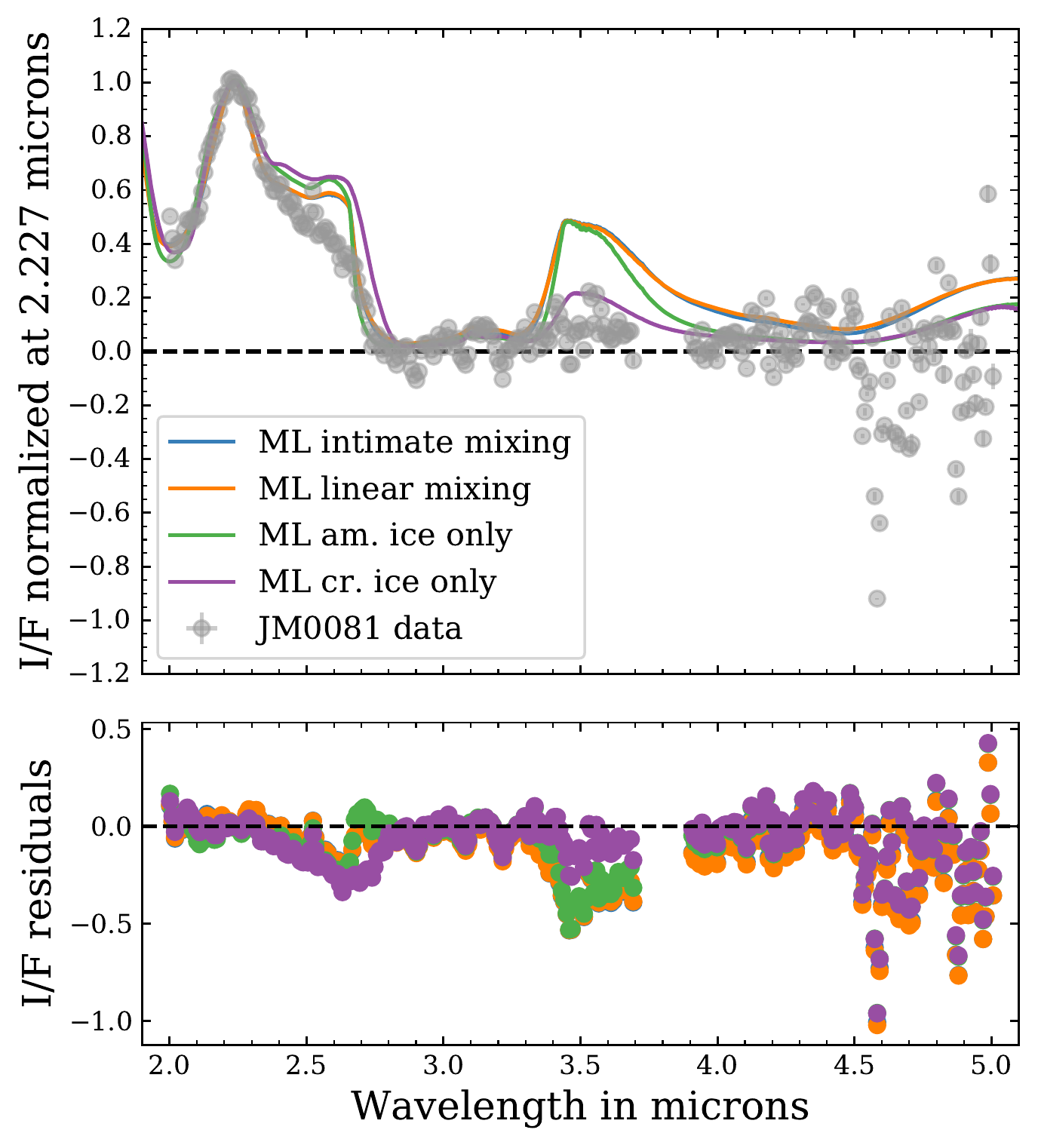}
    \caption{Top panel shows the maximum-likelihood (ML) spectra for the intimate mixing model (blue), the linear mixing model (orange), the amorphous ice only model (green) and the crystalline ice only model (purple), corresponding to the parameter distributions shown in Figure~ \ref{fig:jiram_im}. The blue and orange plots coincide almost perfectly in this figure. The bottom panel shows the residuals for the fits of the model spectra to the JIRAM data, which is shown as grey dots in the top figure. The ML parameters for the intimate mixing model are: log$_{10}D_{am}$ = 1.36, log$_{10}D_{cr}$ = 2.75, $f_{am}$ = 0.9995 and $\phi$ = 0.01. The ML parameters for the linear mixing model are: log$_{10}D_{am}$ = 1.19, log$_{10}D_{cr}$ = 2.85, $f_{am}$ = 0.43 and $\phi$ = 0.01. The ML parameters for the amorphous ice only model are log$_{10}D_{am}$ = 1.64 and $\phi$ = 0.01. The ML parameters for the crystalline ice only model are log$_{10}D_{am}$ = 1.52 and $\phi$ = 0.01. The ML intimate mixing model spectrum has the lowest reduced chi-squared value (Table \ref{Tab: jiram_results}), despite not fitting the 3.6 $\mu$m region as well as the crystalline-ice-only model. This is because it fits the 2.5 $\mu$m region better than the other models, and since the SNR in 2-3 microns $\mu$m region is very high, the shorter wavelength data ends up dictating the overall fit.}
    \label{fig:jiram_im_lm_ML}
\end{figure}

The posterior distributions for our analysis with the two mixing models are shown in Figures~\ref{fig:jiram_im} and \ref{fig:jiram_lm}, and it is evident that the parameters are very sharply constrained for both. While both sets of posteriors have a population of small amorphous ice grains of around 10-20 microns ($23.12_{-1.01}^{+1.01}$ microns for IM and $15.61_{-1.01}^{+1.01}$ for LM), plus a population of large crystalline ice grains of around 500-700 microns ($565.34_{-1.02}^{+1.02}$ microns for IM and $719.94_{-1.03}^{+1.03}$ for LM), the number density fraction of amorphous ice, $f_{am}$, is very high for the IM case, sharply peaked at 0.9995 (equivalent to a mass density fraction, or mass fraction per unit volume, of 0.13 as per eq. \ref{eq:mass_frac}). For the LM case on the other hand, the median $f_{am}$ value is 0.44 (equivalent to an extremely low mass density fraction of around 8E-06).

However, the Bayes factor for the IM model v/s the LM model favours the former, as shown in Table \ref{Tab: jiram_results}. Table \ref{Tab: jiram_results} also shows the Bayes factors for the IM model v/s the single component amorphous-ice-only and crystalline-ice-only models, which are also very strongly in favour the IM model. This means that, within the framework of the considered models, there is evidence for both amorphous ice and crystalline ice in the spectrum. Also shown are the reduced chi-squared values of the best-fitting spectrum of the four models, which also indicate the IM model is the most preferred. We must note that the Bayes factor of the intimate mixture model v/s the other three models is very large, primarily because the data has a very high SNR in the 2-3 $\mu$m wavelength region (see Figure~ \ref{fig:data}). Hence, even a slightly better fit would increase its likelihood value (eq. \ref{eq:bayes_theorem}) and hence the Bayesian evidence by a significant amount. In the same way, a poorer fitting model will be penalized severely with a very low Bayesian evidence. This results in very large Bayes factors when comparing models. The high SNR of the data would also explain why the reduced chi-squared values of the best-fit/ML solutions ($\chi^2_{r,bf}$) are so high, mostly due to the mismatch between the models and the data around 2.5 and 3.6 $\mu$m (see Figure~ \ref{fig:jiram_im_lm_ML}). Its worth noting that in Figure \ref{fig:jiram_im_lm_ML}, the best-fit/ML crystalline-ice-only model spectrum fits the 3.6 $\mu$m region the best. However, its still has a lesser $\chi^2_{r,bf}$ as compared to the ML intimate mixing model spectrum (Table \ref{Tab: jiram_results}) because its fit in the 2.5 $\mu$m region is poorer. Since the 2-3 $\mu$m region has a much higher SNR as compared to the longer wavelength region, it dictates the fits, making the intimate-mixing model the overall preferred one.

\begin{table}
\captionsetup{width=\textwidth}
\centering
\caption{Table \ref{Tab: jiram_results}: Bayesian inference results of different models for fitting the JIRAM data. The columns are same as those in Table \ref{Tab: syn_results}.}
\begin{tabular}{c c c c c c} 
  \hline
  \textbf{Model} & \textbf{Free params} & \textbf{Prior bounds} &\textbf{Solution}  & \textbf{ln$(\boldsymbol{\B_{0i}})$} & $\boldsymbol{\chi^2_{r, bf}}$\\
  \hline
  \makecell{Amorphous \& Crystalline ice \\ (intimate mixing)} & \makecell{log$_{10} D_{am}$, \\ log$_{10} D_{cr}$, \\ $f_{am}$, \\ $\phi$}  & \makecell{(1.0,3.0), \\ (1.0,3.0), \\ (0,1), \\ (0.01,0.52)} & \makecell{1.3640$_{-0.0066}^{+0.0069}$ \vspace{5pt} \\ 2.7523$_{-0.0083}^{+0.0081}$ \vspace{5pt}\\ 0.9995$_{-0.00001}^{+0.00001}$ \vspace{5pt}\\ 0.0126$_{-0.0019}^{+0.0046}$ \vspace{5pt}} & ref. & 65.9 \\
  \hline
  \makecell{Amorphous \& Crystalline ice \\ (linear mixing)} & \makecell{log$_{10} D_{am}$,\\ log$_{10} D_{cr}$, \\ $f_{am}$, \\ $\phi$}  & \makecell{(1.0,3.0), \\ (1.0,3.0), \\ (0,1),\\ (0.01,0.52)} & \makecell{1.1938$_{-0.0085}^{+0.0085}$ \vspace{5pt} \\ 2.8570$_{-0.0116}^{+0.0117}$ \vspace{5pt} \\ 0.4389$_{-0.0032}^{+0.0032}$ \vspace{5pt} \\ 0.0128$_{-0.0021}^{+0.0046}$ \vspace{5pt}} & 335.79 (26.04 $\sigma$) & 68.1\\
  \hline
  Amorphous ice only & \makecell{log$_{10} D_{am}$, \\ $\phi$}  & \makecell{(1.0,3.0),\\ (0.01,0.52)} & \makecell{1.6479$_{-0.0016}^{+0.0016}$ \vspace{5pt} \\ 0.0101$_{-0.0001}^{+0.0002}$ \vspace{5pt}} & 6226.70 (> 30 $\sigma$) & 106.0\\
  \hline
  Crystalline ice only & \makecell{log$_{10} D_{cr}$, \\ $\phi$}  & \makecell{(1.0,3.0), \\ (0.01,0.52)} & \makecell{1.5269$_{-0.0015}^{+0.0015}$ \vspace{5pt} \\ 0.0101$_{-0.0001}^{+0.0002}$ \vspace{5pt}} & 8347.53 (> 30 $\sigma$)& 119.0 \\
  \hline
\end{tabular}
\label{Tab: jiram_results}
\end{table}

\begin{figure}[pos=htbp!]
    \centering
    \includegraphics[width=0.5\linewidth]{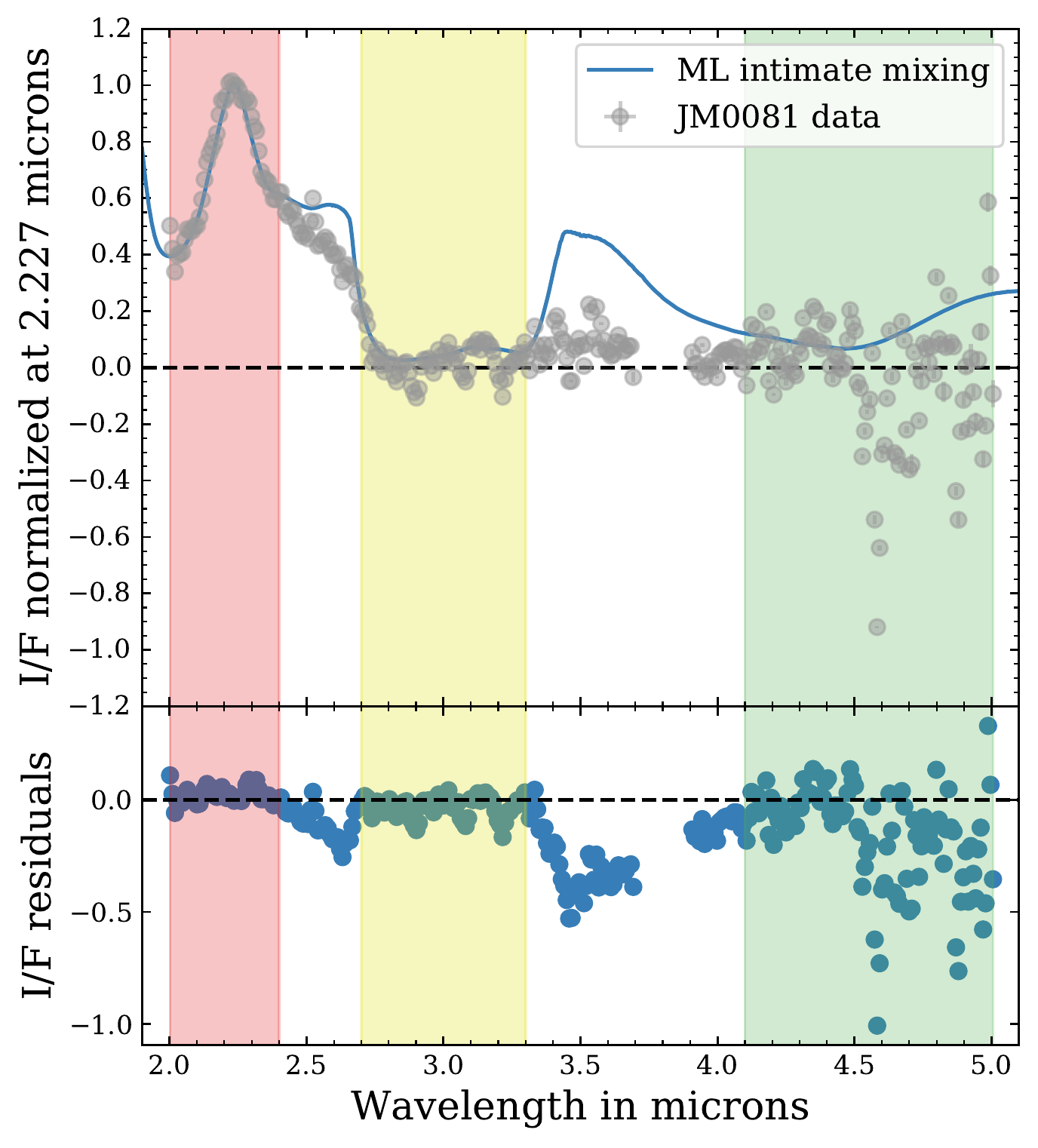}
    \caption{\textit{Top:} The maximum-likelihood model for the two-component intimate mixing model case plotted against the JIRAM data, similar to Figure~ \ref{fig:jiram_im_lm_ML}. The red, yellow and green sections are regions where the model is fitting the data well. \textit{Bottom:} Residuals of the model's fit to the data.}
    \label{fig:sdnr_sections}
\end{figure}

\begin{figure}[pos=htbp!]
    \centering
    \includegraphics[width=0.4\linewidth]{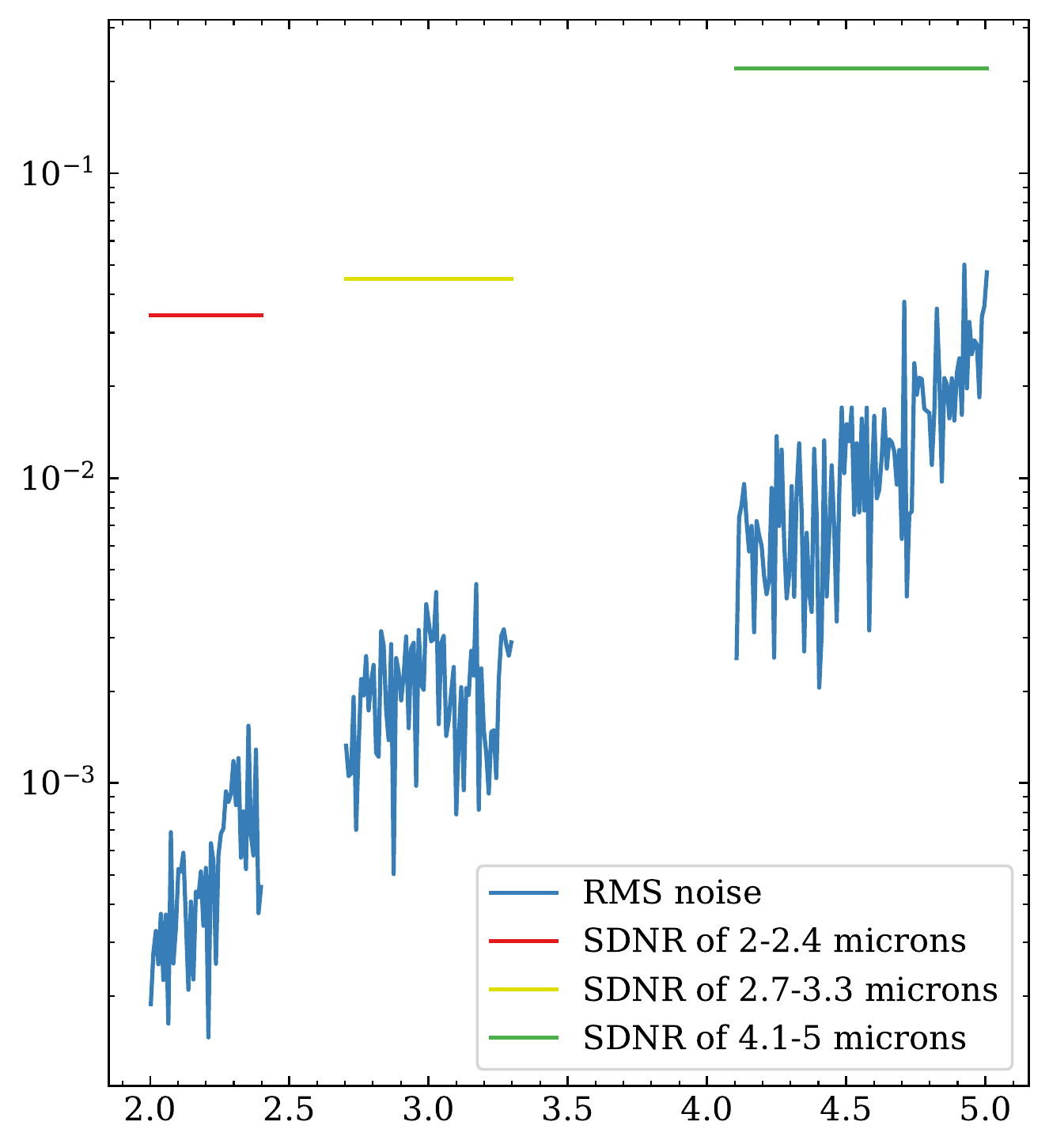}
    \caption{RMS noise of the mean JM0081\_170901\_105708 data is shown (blue) in the wavelength regions corresponding to the red, yellow and green regions from Figure~ \ref{fig:sdnr_sections}. The three horizontal bars correspond to the standard deviation of residuals (referred to as standard deviation of normalized residuals or SDNR), from Figure~ \ref{fig:sdnr_sections}, of the three wavelength regions.}
    \label{fig:rms_vs_sdnr}
\end{figure}

We tested the sensitivity of our results to the estimate of the magnitude of error/noise of the data, by inflating/scaling the noise (error-bar at each data point multiplied by a constant factor) and hence decreasing the SNR. The residuals from the best-fitting/maximum-likelihood solution for the intimate-mixing model (Figure~ \ref{fig:jiram_im_lm_ML}) show a scatter that is larger than our estimate of the RMS noise from JIRAM's NESR (eq. \ref{eq:rms_noise}). This discrepancy is indicative of underestimated noise or missing model features. We estimate a scaling factor for the RMS noise by comparing it to the scatter in the residuals in Figure~ \ref{fig:jiram_im_lm_ML}. Figure~ \ref{fig:sdnr_sections}, just like Figure~ \ref{fig:jiram_im_lm_ML}, shows the residuals of the maximum-likelihood solution of the intimate-mixing model and highlights the sections where the fit is reasonably good. We next compare the standard deviations of the residuals in these wavelength regions to the RMS noise of the JIRAM data, as shown in Figure~ \ref{fig:rms_vs_sdnr}. We can see that the standard deviation of these residuals is roughly greater than the noise we have been using by factors between 20-70. Keeping this range in mind, we performed retrieval analyses for three cases, where the noise is scaled by factors of 20, 30 and 50, whose results are presented in Table \ref{Tab: noise_inflation}. Firstly, we see that the parameter estimations for all four models in all three cases are almost exactly the same, and are also very close to the results from the analysis with the original, un-scaled noise as presented in Table \ref{Tab: jiram_results}. This highlights the robustness of our inference analysis results. Secondly, we see that the Bayes factors and the $\sigma$-significance for the preference of the intimate-mixing model get lower as we increase the noise from Case-1 to Case-3. For Case-3 (with 50x noise) we find that the intimate-mixing model is still preferred over the linear-mixing model at 3.9$\sigma$ confidence. We also find that crystalline ice is detected at a 22.0$\sigma$ confidence (or in other words, the intimate-mixing model is preferred over the amorphous-ice only model at 22.0$\sigma$ confidence) and amorphous ice is detected at a 25.6$\sigma$ confidence. Finally, we see that the reduced chi-squared values also decrease, approaching unity with increased noise inflation. The chi-square values are still relatively large due to the mismatch in the 2.5 and 3.6 $\mu$m wavelength regions between the data and our ML intimate mixing model, as can be seen in Figure~ \ref{fig:jiram_im_lm_ML}. 

We also experimented with higher values of the noise-scaling factor, and found that for a noise-scaling factor of 80 and higher, the difference between the Bayesian evidence for intimate mixing and linear mixing models is not significant. This decrease in evidence of the intimate-mixing model over the linear-mixing model is expected because for very noisy data, it is difficult for the Bayesian inference framework to differentiate between complex models.

\begin{table}
\centering
\caption{Table \ref{Tab: noise_inflation}: Bayesian inference results of different models for fitting JIRAM data, same as Table \ref{Tab: jiram_results}, but for three different cases. Case-1, Case-2, and Case-3 involve analyzing the mean JIRAM data with the RMS noise inflated by factors of 20, 30 and 50, respectively.}
 \resizebox{0.75\textwidth}{!}{\begin{minipage}{\textwidth}
\hspace*{-3cm}\begin{tabular}{p{2cm} p{2cm} p{2.5cm} p{1cm} p{2cm} p{2.5cm} p{1cm} p{2cm} p{2.5cm} p{1cm}}
  \hline
  \textbf{Model} & \makecell{\textbf{Case-1:} \\ \textbf{ML solution}}  &\textbf{ Case-1: ln$(\boldsymbol{\B_{0,i}})$} & \textbf{Case-1: }$\boldsymbol{\chi^2_{reduced}}$ & \textbf{Case-2: ML solution}  & \textbf{Case-2: ln$(\boldsymbol{\B_{0,i}})$} & \textbf{Case-2: }$\boldsymbol{\chi^2_{reduced}}$ & \textbf{Case-3:  ML solution}  & \textbf{Case-3: ln$(\boldsymbol{\B_{0,i}})$} & \textbf{Case-3: }$\boldsymbol{\chi^2_{reduced}}$\\
  \hline
 \makecell{Amorphous  \\\& Crystalline ice \\ (intimate mixing)}  & \makecell{1.36, 2.75, \\ 0.9995, 0.01} & ref. & 16.5 & \makecell{1.36, 2.75, \\ 0.9995, 0.01} & ref. & 7.32 & \makecell{1.36, 2.74, \\ 0.9995, 0.01} & ref. & 2.64  \\
  \hline
 \makecell{Amorphous \\ \& Crystalline ice \\ (linear mixing)} & \makecell{1.19, 2.85, \\ 0.44, 0.01} & 76.69 (12.59 $\sigma$) & 17.0 & \makecell{1.19, 2.86, \\ 0.44, 0.01} & 28.59 (7.84 $\sigma$) & 7.57 & \makecell{1.19, 2.85, \\ 0.44, 0.01} & 6.32 (3.98 $\sigma$) & 2.73\\
  \hline
  Amorphous ice only &  1.65, 0.01 & 546.85 (> 30$\sigma$) & 26.4 & 1.65, 0.01 & 680.36 (36.98 $\sigma$) & 11.7 & 1.64, 0.01 & 240.05 (22.05 $\sigma$) & 4.23\\
  \hline
  Crystalline ice only  & 1.52, 0.01 & 2077.08 (> 30$\sigma$) & 29.9 & 1.52, 0.01 & 916.05 (> 30$\sigma$) & 13.3 & 1.52, 0.01 & 324.38 (25.60 $\sigma$) & 4.78 \\
  \hline
 
\end{tabular}
\label{Tab: noise_inflation}
\end{minipage}}
\end{table}

Finally, a common result from all the analyses we have performed is that the retrieved values for the filling factor $\phi$ are close to the lower limit of the bound we had set for it. This is also reflected in the probability distribution of $\phi$, in Figure~s \ref{fig:jiram_im} and \ref{fig:jiram_lm}, that peak at around 0.01. This indicates a very porous regolith, which is not in agreement with the polarimetric studies of Europa by \citet{poch_polarimetry_2018}, who suggest that Europa is possibly covered by sintered grains, leading to a more compact regolith and hence a larger value for filling factor $\phi$. We suspect that the low value of $\phi$ we derive suffers from the normalization effect on the I/F spectra we have used. Normalized Hapke model spectra are very mildly sensitive to $\phi$ (Figure~ \ref{fig:sensitivity}). The major effect of $\phi$ on the model spectrum is changing the absolute reflectance (Figure~ \ref{fig:phi_spectra}), which is removed as we are working with normalized spectra.\\

\section{Discussion and Conclusions} \label{sec:disc_and_conc}

\begin{figure}[pos=htbp!]
    \centering
    \includegraphics[width=0.75\linewidth]{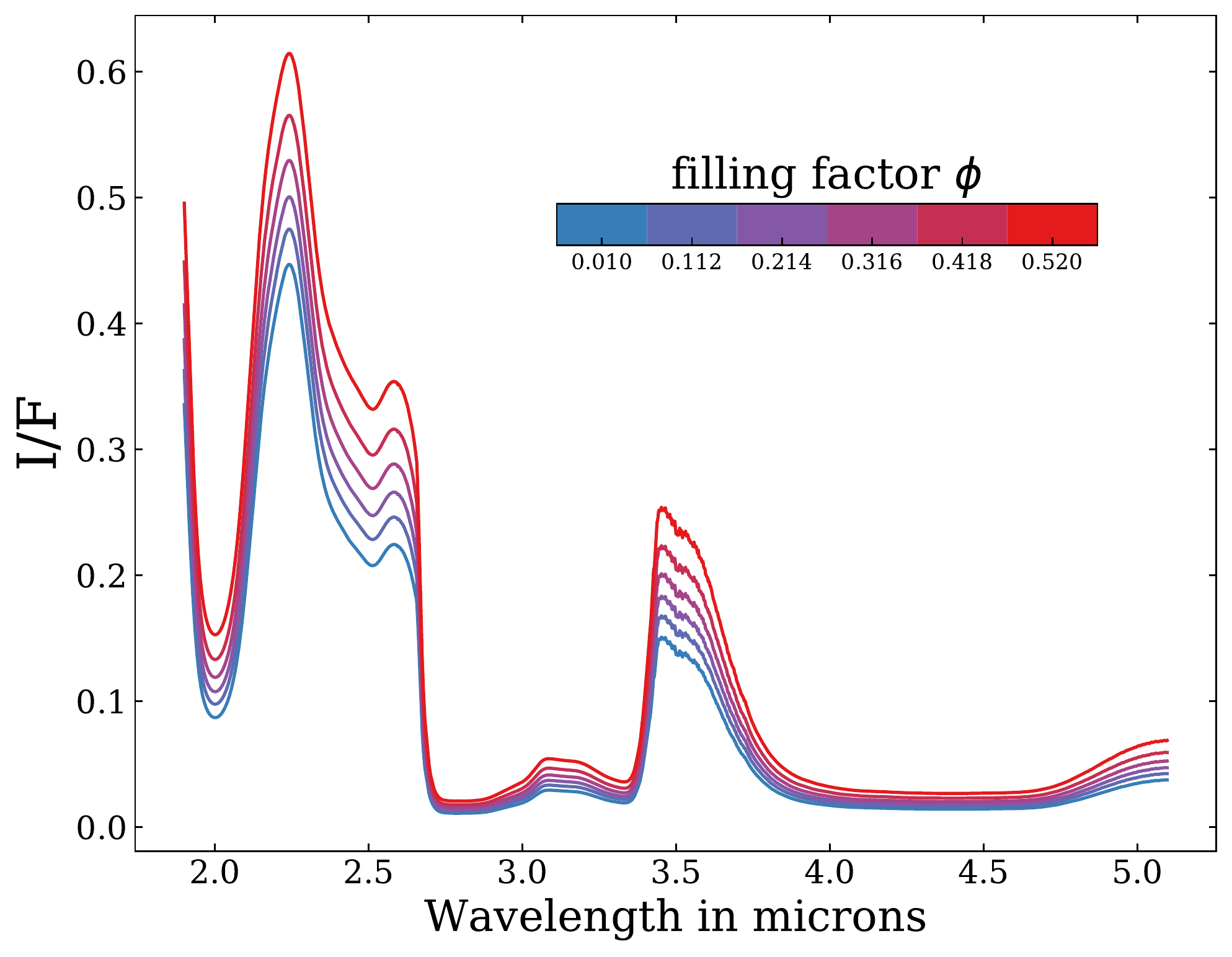}
    \caption{Effect of the filling factor $\phi$ on the absolute reflectance of a model spectrum of amorphous ice grains of average grain size $D_{am} = 200 \ \mu$m. Observation geometry parameters of the model are same as the JIRAM data (Table \ref{Tab: data}): incidence angle $= 28.5$ deg., emission angle $= 73.4$ deg. and phase angle $g = 91.5$ deg.}
    \label{fig:phi_spectra}
\end{figure}

\begin{figure}[pos=htbp!]
    \centering
    \includegraphics[width=0.75\linewidth]{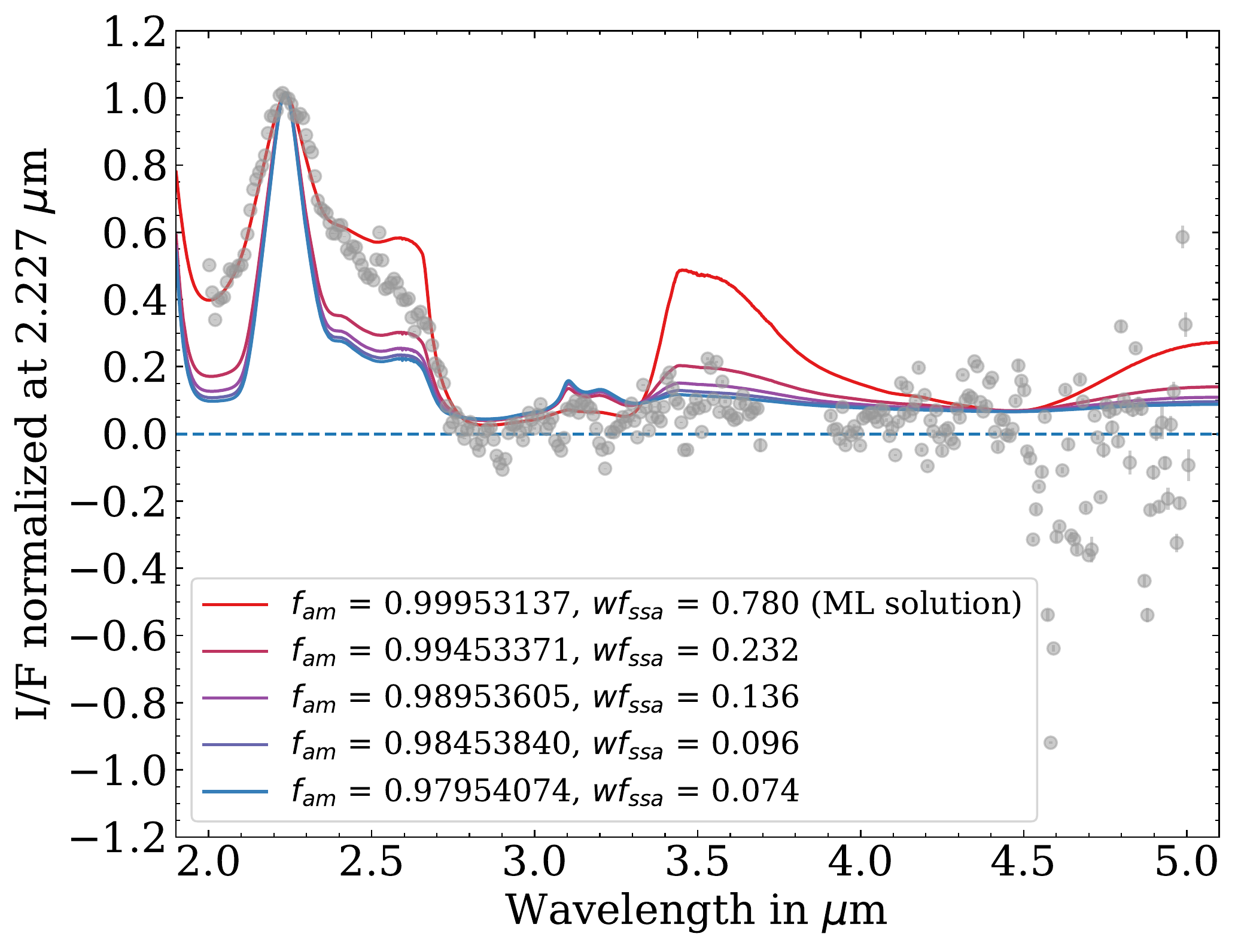}
    \caption{Sensitivity of the maximum-likelihood (ML) solution for the intimate-mixing model's fit to the JIRAM data. The solid lines are for different values of $f_{am}$, the number density fraction of amorphous ice. Mentioned next to the $f_{am}$ values in the legend are the corresponding weight fractions for the single scattering albedo, $wf_{ssa}$ (see the mixing equation, eq. \ref{eq:im_eqn}). The red line corresponds to the ML solution of the intimate-mixing model. The gray scatter points are the JIRAM data.}
    \label{fig:f_am_sensitivity}
\end{figure}

We present a Bayesian retrieval analysis of selected spectra of Europa collected by \textit{Juno}/JIRAM (Figure~ \ref{fig:data}). We validate our analysis framework on simulated and laboratory data in section \ref{sec:synthetic_and_lab}. We consider multiple possible models to explain the Juno/JIRAM observations, including a two-component (amorphous plus crystalline ice) intimate mixing model, a two-component linear mixing model, a single-component amorphous ice model and a single component crystalline ice model.  Our main results are summarized in Table \ref{Tab: jiram_results}. Our key findings are the following:  

\begin{enumerate}
    \item The two-component intimate-mixing (TC-IM) model is preferred over the linear-mixing model (to 26$\sigma$ confidence), with tightly constrained parameters (Figure~\ref{fig:jiram_im}). The retrieved parameter posterior probability distributions, as shown in Figure~\ref{fig:jiram_im}, correspond to a mixture with a very large number density fraction ($\approx$ 0.9995) of small ($\approx$ 20 microns) amorphous ice grains. Since the linear-mixing model is not favored by our Bayesian model comparisons, the surface we are observing is better interpreted as an intimate mixture of large and small grains instead of discrete patches.
    \item The very high fraction of small amorphous-ice grains can be understood by considering the sensitivity of the maximum-likelihood (ML) solution of the intimate-mixing model to the number density fraction of amorphous ice, or $f_{am}$. This extreme sensitivity is demonstrated in Figure~ \ref{fig:f_am_sensitivity}, where we vary $f_{am}$ by tiny amounts and inspect its effect of the spectrum. Starting with the ML value of $f_{am} \approx$ 0.9995 (keeping all other parameters constant), we see that even a 0.5\% change in $f_{am}$ causes the model spectrum to change significantly and degrade its fit to the 2-2.7 $\mu$m region data. While the fit actually ends up improving in the region around 3.5 $\mu$m, the fit to the data in the 2-2.7 $\mu$m region is the deciding factor due to its very high SNR  (see Figure \ref{fig:data}). Hence values of $f_{am}$ even slightly lower than the ML value of 0.9995 are not favoured by the Bayesian framework. The orders of magnitude difference in SNR between the 2-2.7 $\mu$m region data and the 3.0-3.5 $\mu$m region data also explains why the maximum-likelihood solution of the TC-IM model is favored over the crystalline-ice-only model in our Bayesian analysis, despite the fact that the latter fits the 3-3.5 $\mu$m region data better, as shown in Figure \ref{fig:jiram_im_lm_ML}.
    \item The TC-IM model's high sensitivity to $f_{am}$ stems from the influence of the abundance or number-density fractions on the average single scattering albedo (SSA) of a mixture, defined by the mixing eq. \ref{eq:im_eqn}. In eq. \ref{eq:im_eqn}, the `weight' of each mixture component depends on the product of its number density fraction and square of its grain-size. In our ML solution, the amorphous ice grain size is an order of magnitude smaller than that of crystalline ice ($\approx$20 microns v/s $\approx$560 microns). Given the small size of the amorphous ice grains, a very high number-density fraction ($f_{am}$) is needed for amorphous ice's SSA to have a significant weight in eq. \ref{eq:im_eqn}. In the legend of Figure~ \ref{fig:f_am_sensitivity}, along with different values of $f_{am}$, their corresponding weight-fraction in eq. \ref{eq:im_eqn} (denoted here by $wf_{ssa}$) are also mentioned. As can be noted, even though $f_{am}$ decreases marginally, the $wf_{ssa}$ values decrease by over one order of magnitude.
    \item Figure~\ref{fig:jiram_im_lm_ML} shows that the ML solution of the intimate-mixing model is not able to fit the data at around 2.5 and 3.6 $\mu$m, which might indicate the presence of non-ice components that strongly absorb in these wavelength regions. For the 3.6 $\mu$m region, the strong absorption could also be an indication of the fact that a more complex grain size distribution may be needed, which is beyond the scope of the model we have used. 
    \item The porosity factor distribution peaks at the lower limit of the prior range 0.01 for all the cases, indicating a very porous regolith. However, we should be careful about this conclusion since the major effect of $\phi$ on the model spectrum is changing the absolute reflectance (Figure~ \ref{fig:phi_spectra}), which is removed as we are working with a normalized spectrum.   
\end{enumerate}

The overabundance of amorphous ice grains we are finding is not surprising given that the environmental conditions necessary to maintain water ice in both crystalline and amorphous states exist on the surface of Europa \citep{hansen_amorphous_2004,carlson_europas_2009}. It is well established that a water ice layer undergoes lattice structure transition, converting crystalline ice to amorphous ice, through impacts of high energy electrons and ions from Jupiter's magnetosphere \citep{baratta_31_1991,strazzulla_ion-beam-induced_1992,moore_far-infrared_1992,hudson_far-ir_1995,leto_structural_1996,leto_ly-alpha_2003,leto_reflectance_2005} and from condensation of sublimated and sputtered molecules \citep{baragiola_water_2003}. \citet{filacchione_serendipitous_2019}, who published JIRAM's observations of Europa, also presented an amorphous v/s crystalline ice composition analysis based on the shift of the 2 $\mu$m absorption band center and the value of the ratio I/F$_{(3.100 \ \mu m)}$~/ I/F$_{(2.847 \ \mu m)}$. They find that the JM0081 set of spectra, to which the data used in this work belongs to (see Table \ref{Tab: data}), has a larger presence of amorphous ice grains. They also find that the northern hemisphere in general, where our data are located, has more amorphization. Electron flux calculations done by \citet{nordheim_preservation_2018} showed that there is significant radiation processing of the surface material even in mid-to high latitudes of Europa, where our data comes from (20.6-24$\degree$ N latitude and 37.4-40.8$\degree$ W longitude). This further supports the prevalence of amorphous ice grains in the region we are observing. Finally, its important to keep in mind that water ice's optical skin depth is very shallow in the spectral range being probed by the JIRAM data (2-5 $\mu$m). We are observing the top-most millimeter to sub-micron layer.  \citet{hansen_amorphous_2004} analyzed the 3.1 $\mu$m Fresnel reflection band in the Galileo/NIMS data of Europa, which is diagnostic of the lattice order (amorphous v/s crystalline) of water ice in the top micrometers of the surface, and found it to be predominantly amorphous, which is what we find as well. Studying a broader or a different wavelength range, for example 1-2 $\mu$m, where photons are penetrating deeper into the surface, could allow to probe the composition at deeper scales.

The small grain sizes of amorphous ice that we have retrieved are consistent with a strong correlation of grain-sizes of water-ice with sputtering rates of different locations on Europa \citep{clark_frost_1983, cassidy_magnetospheric_2013}. Larger sputtering rates are thought to produce regolith with predominantly larger grain-sizes, as smaller grains are destroyed faster. Our data comes from the leading hemisphere, which has lower sputtering or ion erosion rates \citep{cassidy_magnetospheric_2013} and hence one would expect water-ice to be predominantly of smaller grain-size, which is consistent with our findings. Our result is also consistent with previous analyses of spectroscopic data from the leading hemisphere of Europa, where the calculated water-ice grain-sizes are on the order of 10s of microns \citep[e.g.][]{shirley_europas_2010, dalton_europas_2012}. As a final note, we acknowledge that our analysis was fundamentally limited by the normalization of the \textit{Juno}/JIRAM spectra, due to lack of the full photometric information needed to calculate the absolute $I/F$ values. To resolve this it is necessary to observe the same area of a surface at different illumination/viewing geometries which allows us to, using Hapke's model, disentangle completely photometric effects from effects due to the composition (endmembers, mixing fractions) and physical state of the surface (grain size, roughness, porosity, temperature, etc.). With the current JIRAM dataset this is not feasible. \\ 

\noindent \emph{Future work} \\

The mismatch between our best model solution and the observations around 2.5 and 3.6 $\mu$m potentially indicates the presence of non-ice components not currently included in our model. Our new technique therefore holds the promise of being able to identify and provide detection significances for these species in future work and constrain their abundances and physical properties. A number of non-water-ice species have been proposed to exist on Europa’s surface (e.g. \citet{carlson_europas_2009,trumbo_sodium_2019}). Future work will include incorporating these species into the analysis of this new and rich Juno/JIRAM data. The most significant bottle-neck of any spectroscopic analysis is the availability of optical constants or refractive indices of the components one wishes to include in the model. Although the scope of this work was to explore the water-ice composition of the JIRAM data, in terms of the distribution of amorphous v/s crystalline ice, a more comprehensive analysis is still limited by the non-availability of relevant optical constants of other components (like hydrated salts and acids) that have been detected on Europa. Our work, like many previous studies of Europan data, highlights the need for laboratory measurements of optical constants in the 2-5 $\mu$m wavelength regime and in the right temperature regime of 80-130 K. A rich database of NIR optical constants is needed to analyze the wealth of NIR Europan data that exists \citep{filacchione_serendipitous_2019, hansen_widespread_2008, mccord_cassini_2004, grundy_new_2007, hand_keck_2013, ligier_vlt/sinfoni_2016} and build a comprehensive picture of Europa’s surface.

\section{Acknowledgements}

We would like to thank Roger Clark for providing us the laboratory spectrum that was vital for validating our analysis methodology, Alessandro Mura for his useful comments and insights regarding the pre-processing of JIRAM data and Will Grundy and Stephen Tegler for frutiful discussions regarding our results in the early stages of the project. Jonathan Lunine acknowledges the financial support of \textit{Juno} mission subcontract D99069MO from the Southwest Research Institute. Gianrico Filacchione and Mauro Ciarniello acknowledge the financial support of an INAF grant (INAF grant (Call a sostegno dei progetti Mainstream INAF, PI: G. Filacchione, Funzione Obiettivo 1.05.01.86.11). 

\section{Data Availability}

The raw radiance files corresponding to all observations from the JIRAM instrument on \textit{Juno} can be found on NASA's Planetary Data System, available at \url{https://pds-atmospheres.nmsu.edu/data_and_services/atmospheres_data/JUNO/jiram.html}.

\bibliographystyle{cas-model2-names}

\bibliography{manuscript}


\newpage

\appendix

\section{Hapke equation parameters}

Here we describe in detail various parameters that appear in the Hapke RT equation (eq. \ref{eq:hapke_RT}). \\

\noindent \emph{The porosity coefficient $K$} \\

\noindent \citet{hapke_bidirectional_2008,hapke_theory_2012} added a parameter to the previous iteration of their model, that accounts for the dependence of bidirectional reflectance on the porosity of the regolith. It has been found that compression of a loosely packed powder increases its reflectance, albeit the margin by which the reflectance increases with greater compaction diminishes for progressively higher albedo materials \citep[e.g.][]{hapke_bidirectional_2008,helfenstein_testing_2011}. Hence, a parameter $K$, the \textit{porosity coefficient}, features in equation 7. It is directly related to the filling factor of the material (i.e., the total fraction of volume occupied by particles), $\phi$, via the equation

\begin{gather} \label{eq:K_defn}
    K = \dfrac{\textrm{-ln}(1 - 1.209\phi^{2/3})}{1.209\phi^{2/3}}
\end{gather}
The filling factor $\Phi$ for a mixture with $j$ components is defined as

\begin{gather}
    \Phi = \sum_j N_j v_j = \sum_j N_j \frac{4}{3} \pi \left(\frac{D_j}{2}\right)^3
\end{gather}
where $N_j$ is the number density (units m$^{-3}$) and $v_j$ is the volume of a single particle/grain of component $j$ of diameter $D_j$.
Under the assumption that particles are sparsely packed, we can put $K=1.0$ and equation 7 reduces to its most used version (eg. \citet{carlson_distribution_2005,ciarniello_hapke_2011,clark_surface_2012}).  Putting K=1 assumes a "fluffy" surface which seems more reasonable for Saturn's satellites \citep{clark_surface_2012, ciarniello_hapke_2011} where Enceladus' plumes particles are coating the surfaces of many satellites, but it is less valid for Europa whose surface is dominated by large ($> 100$ microns) grains \citep[e.g.][]{cassidy_magnetospheric_2013, filacchione_serendipitous_2019}. The higher diurnal temperature of Europa allows sintering of small ice grains in larger ones. The compactness of the Europa surface (corresponding to $K>1$) has been confirmed by polarimetric observations \citep{poch_polarimetry_2018}.\\
 
\noindent \emph{The single-scattering albedo $\omega$} \\
 
\noindent In equation \ref{eq:hapke_RT}, the dependence of the reflectance $I/F$ on the optical constants of the constituent(s) comes from the single-scattering albedo $\omega$, which is a function of the material's optical constants ($n$ and $k$) and grain diameter $D$. As defined by \citet{hapke_bidirectional_1981,hapke_theory_2012}, for a medium that consists of only one type of particle that is large compared to the wavelength, $\omega$ is given by
 
 \begin{gather}
     \omega = \dfrac{N\sigma Q_S}{N\sigma Q_E} = \dfrac{Q_S}{Q_E} = Q_s
 \end{gather}
where $Q_S$ is the \textit{volume-average scattering efficiency including diffraction}, $Q_s$ is the \textit{volume-average scattering efficiency excluding diffraction} and $Q_E$ is the \textit{volume-average extinction efficiency}. Here, we assume that $Q_E=1$ and $Q_S = Q_s$ for large particles in a medium in which the particles are in contact. In the equivalent-slab approximation, as described in \citet{hapke_theory_2012}, the single-scattering albedo is given by

\begin{gather} \label{eq:ssa}
    \omega = Q_s = S_e + (1 - S_e) \dfrac{(1 - S_i)\Theta}{1 - S_i\Theta}
\end{gather}
where $S_e$ and $S_i$ are, respectively, the average \textit{Fresnal reflection coefficients for externally and internally incident light}. In the equivalent-slab approximation, they are given by
\begin{gather}
    S_e = 0.0587 + 0.8543 R(0) + 0.0870 R(0)^2 \\
    S_i \approx 1 - \dfrac{1}{n}[0.9413 - 0.8543 R(0) - 0.0870 R(0)^2]
\end{gather}
where $n$ is the real part of refractive index of the material and  $R(0)$ is the \textit{normal specular reflection coefficient} given by
\begin{gather}
    R(0) = \dfrac{(n - 1)^2 + k^2}{(n + 1)^2 + k^2}
\end{gather}
where $k$ is the imaginary part of the refractive index. $\Theta$ is the particle \textit{internal transmission factor}, given by

\begin{gather} \label{eq:in_trans_factor}
    \Theta = \dfrac{r_i + \textrm{exp}(-\sqrt{\alpha(\alpha + s)}\langle D \rangle)}{1 + r_i \textrm{exp}(-\sqrt{\alpha(\alpha + s)}\langle D \rangle)}
\end{gather}
where $r_i = \frac{1 - \sqrt{\frac{\alpha}{\alpha + s}}}{1 + \sqrt{\frac{\alpha}{\alpha + s}}}$ is the \textit{internal diffusive reflectance}, $\alpha$ is the \textit{absorption coefficient} ($= 4\pi k/\lambda$), $s$ is the \textit{internal scattering coefficient inside the particle} and $\langle D \rangle$ is the \textit{effective particle size} in the equivalent-slab approximation defined as

\begin{gather} \label{eq:<D>}
    \langle D \rangle = \dfrac{2}{3} \Big[ n^2 - \dfrac{1}{n}(n^2 - 1)^{3/2}\Big]D
\end{gather}

\medskip

\noindent \emph{The phase function $P$} \\

\noindent We use the two-parameter \textit{Henyey-Greenstein} phase function \citep{henyey_diffuse_1941}

\begin{gather}
    P = \dfrac{1+c}{2}\dfrac{1 - b^2}{(1 - 2b\textrm{cos}g + b^2)^{3/2}} + \dfrac{1-c}{2}\dfrac{1 - b^2}{(1 + 2b\textrm{cos}g + b^2)^{3/2}}
\end{gather}
The first term in the RHS describes the backward lobe/ scattering  whereas the second terms describes the forward lobe/scattering of the particle. The $b$ parameter is constrained to lie in the range $0 \leq b \leq 1$; there is no constraint on c except that $P(g) \geq 0$ everywhere. In the equivalent-slab approximation we are following, $c$ is simply

\begin{gather}
    c = \dfrac{\Delta Q_s}{Q_s} \\
\end{gather}
$b$ can then be calculated from an empirical relation as

\begin{gather}
    b = 0.15 + \dfrac{0.05}{(1 + \Delta Q_s/Q_s)^{4/3}}    
\end{gather}
where $Q_s$ is defined in eq. 11 and $\Delta Q_s$ is the \textit{scattering efficiency difference}, which is the difference between the back- and forward-scattering efficiency of a slab and is given by 

\begin{gather}
    \Delta Q_s = S_e + (1 - S_e)(1 - S_i)\dfrac{\Psi}{1 - S_i \Psi}
\end{gather}
$\Psi$ is the \textit{scattering efficiency difference factor} given by

\begin{gather}
    \Psi = \dfrac{r_i - \textrm{exp}(-\sqrt{\alpha(\alpha + s)}\langle D \rangle)}{1 - r_i \textrm{exp}(-\sqrt{\alpha(\alpha + s)}\langle D \rangle)}
\end{gather}
Although there is also a three-parameter \textit{Henyey-Greenstein} function, which allows for the backward and forward lobes to be modelled more independently, it has been shown that the improvement in fits of phase function for real materials, including water ice, is marginal (eg. \citet{domingue_re-analysis_1997,hartman_scattering_1998}). Moreover, the two-parameter Henyey-Greenstein function has been shown to be representative of a wide variety of planetary regolith, including the icy particles on Europa \citep{hapke_theory_2012}. \\

\noindent \emph{The Ambartsumian-Chandrasekhar $H$ function} \\

\noindent As described in \citet{hapke_theory_2012}, an excellent approximation of the Ambartsumian-Chandrasekhar $H$ function is given by

\begin{gather} \label{eq:chandra}
    H(\omega, x) \approx \dfrac{1}{1 - \omega x\left[r_0 + \frac{1 - 2 r_0 x}{2} \textrm{ln} \Big(\frac{1+x}{x}\Big)\right]}
\end{gather}
where $r_0$ is the \textit{diffusive reflectance} . For isotropic scatterers, it is given by

\begin{gather}
    r_0 = \dfrac{1 - \gamma}{1 + \gamma}
\end{gather}
and $\gamma$ is the \textit{albedo factor} given by

\begin{gather}
    \gamma = (1 - \omega)^{1/2}
\end{gather}
Eq. \ref{eq:chandra} has been derived under the Isotropic Multiple Scattering Approximation or IMSA, which goes back to the earliest version of Hapke's model \citep{hapke_bidirectional_1981}. Through Monte-Carlo ray tracing simulations, \citet{ciarniello_test_2014} have shown that this formulation is satisfactory when modelling porous material at high phase angles, away from the regime of opposition surge (phase angle for our data is 90 \degree).  \\

\end{document}